\documentclass[10pt,final,journal,compsoc,letterpaper,twoside]{IEEEtran}

\ifCLASSOPTIONcompsoc
  \usepackage[nocompress]{cite}
\else
  \usepackage{cite}
\fi
\ifCLASSINFOpdf
  \usepackage[pdftex]{graphicx}
  \graphicspath{{figures/}{pictures/}{images/}{./}}
  \DeclareGraphicsExtensions{.pdf,.jpeg,.png}
\else
\fi
\usepackage{amsmath}
\ifCLASSOPTIONcompsoc
 \usepackage[caption=false,font=footnotesize,labelfont=sf,textfont=sf]{subfig}
\else
  \usepackage[caption=false,font=footnotesize]{subfig}
\fi
\usepackage{tabularx,multirow}
\usepackage{paralist}
\usepackage{hyperref}

\begin{document}
%
\title{On the Readability of Abstract Set Visualizations}
%
%
%
%

\author{Markus~Wallinger,
        Ben~Jacobsen,
        Stephen~Kobourov,
        and~Martin~N\"ollenburg
\IEEEcompsocitemizethanks{\IEEEcompsocthanksitem 
M.~Wallinger and M.~N\"ollenburg are with the Algorithms and Complexity Group, TU Wien, Vienna, Austria
\IEEEcompsocthanksitem B.~Jacobsen and S.~Kobourov are with the Dept. of Computer Science, University of Arizona, Tucson, AZ}
}

%
%

\markboth{}%
{Wallinger \MakeLowercase{\textit{et al.}}: On the Readability of Abstract Set Visualizations}
\IEEEtitleabstractindextext{%
\begin{abstract}
    
Set systems are used to model data that naturally arises in many contexts: social networks have communities, musicians have genres, and patients have symptoms. 
Visualizations that accurately reflect the information in the underlying set system make it possible to identify the set elements, the sets themselves, and the relationships between the sets. 
In static contexts, such as print media or infographics, it is necessary to capture this information without the help of interactions.  
With this in mind, we consider three different systems for medium-sized set data, LineSets, EulerView, and MetroSets, and report the results of a controlled human-subjects experiment comparing their effectiveness.
Specifically, we evaluate the performance, in terms of time and error, on tasks that cover the spectrum of static set-based tasks. 
We also collect and analyze qualitative data about the three different visualization systems. Our results include statistically significant differences, suggesting that MetroSets performs and scales better.


\end{abstract} 

}

\maketitle

\IEEEdisplaynontitleabstractindextext

%
\IEEEpeerreviewmaketitle

\IEEEraisesectionheading{\section{Introduction}\label{sec:introduction}}

%
%
%
%
\IEEEPARstart{S}{et} systems naturally model  
data with categorical attributes that occur frequently in data science and analytics in various application domains. 
Examples are actors in social networks with different (overlapping) community structures they are members of, patients in health care data and the different types of symptoms they show, or artists and bands that belong to various musical genres. 
More generally, a set system $(\mathcal U, \mathcal E)$ is comprised of a universe $\mathcal U$ of \emph{elements} (actors, patients, or artists) that are grouped into different \emph{subsets} $S \subseteq \mathcal U$, which together form a subset family $\mathcal E \subseteq 2^{\mathcal U}$ (communities, symptoms, or genres).
Set systems can also be modeled as \emph{hypergraphs}, where $\mathcal U$ is the vertex set and $\mathcal E$ is the set of \emph{hyperedges}. 
Each hyperedge $S \in \mathcal E$ is a subset $S \subseteq \mathcal U$ of vertices.
Hypergraphs generalize graphs, which have the additional restriction that (hyper-)edges must consist of exactly two vertices.

The most popular visualization style for small set systems are Euler (and Venn) diagrams, showing sets as closed shapes with possible overlaps indicating set relations such as intersection or containment.
However, Euler diagrams do not necessarily show individual elements and do not scale well. 
Therefore many different visualization styles have been proposed in the literature, ranging from overlay techniques for pre-embedded elements in the plane to more scalable node-link and matrix-based techniques. 
A state-of-the-art report by Alsallakh et al.~\cite{alsallakh2014visualizing} includes a taxonomy for classifying the different visualization techniques as well as a collection of 26 general tasks in three categories to be solved using set visualizations. 
One of the observations from their survey is a ``clear lack of empirical user studies that assess the effectiveness of different techniques in performing different tasks''~\cite{alsallakh2014visualizing}.

There are a few previous empirical evaluations. Some  considered spatial set systems with embedded elements~\cite{mhsad-khvt-13}, or focused on different visual parameters in a single type of visualization style such as Euler diagrams~\cite{csrmb-vsecdt-14}
or LineSets~\cite{ahrc-dslnvt-11}.
Others studied specific tasks in combined visualizations of set systems and underlying networks~\cite{DBLP:conf/diagrams/BaimagambetovS020,DBLP:journals/isci/RodgersSAMBT16} or compared  visualization techniques that do not show individual elements such as linear and mosaic diagrams~\cite{lm-clmdv-19}.


The focus of our study is on evaluating the readability of set visualizations for abstract, non-spatial data generated by publicly available visualization systems, by measuring performance of representative element-based and set-based tasks. 
In particular, this means that both individual elements in $\mathcal U$ and all the sets in $\mathcal E$ must be shown and labeled. 
Further, we aimed at visualization techniques that can be intuitively understood by non-experts and thus serve as candidates for set visualizations to be used by data journalists, e.g., in static infographics for print media, or for sharing images about interesting datasets in social media.
This focus of the study limits the size of datasets to less than ten sets and fewer than a hundred  elements. Datasets larger than that require interaction and navigation (e.g., filtering, zooming, panning). 

\begin{figure*}[t]
    \centering
    \subfloat {
		\includegraphics[width=0.33\textwidth]{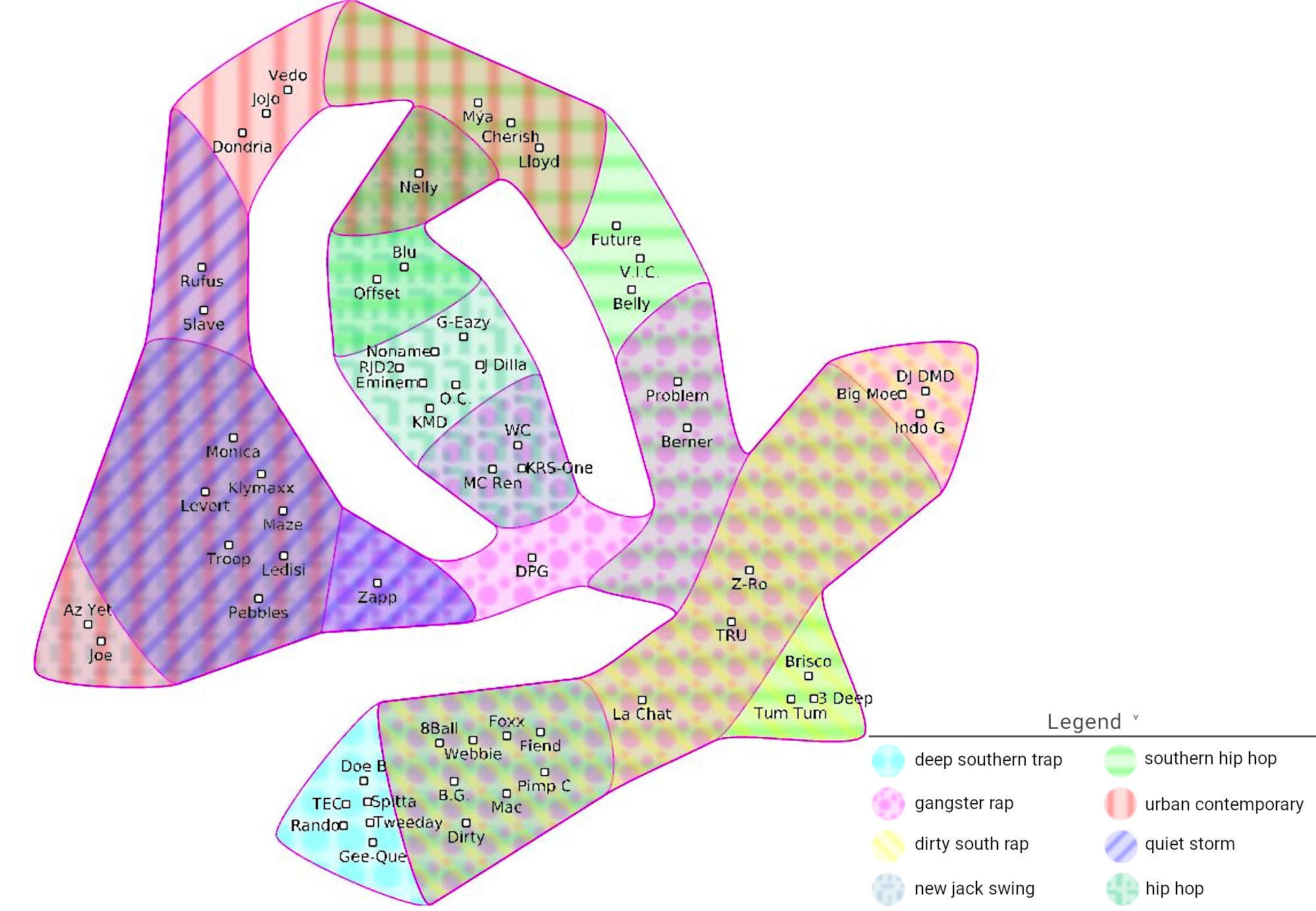}
	}
		\subfloat {
		\includegraphics[width=0.33\textwidth]{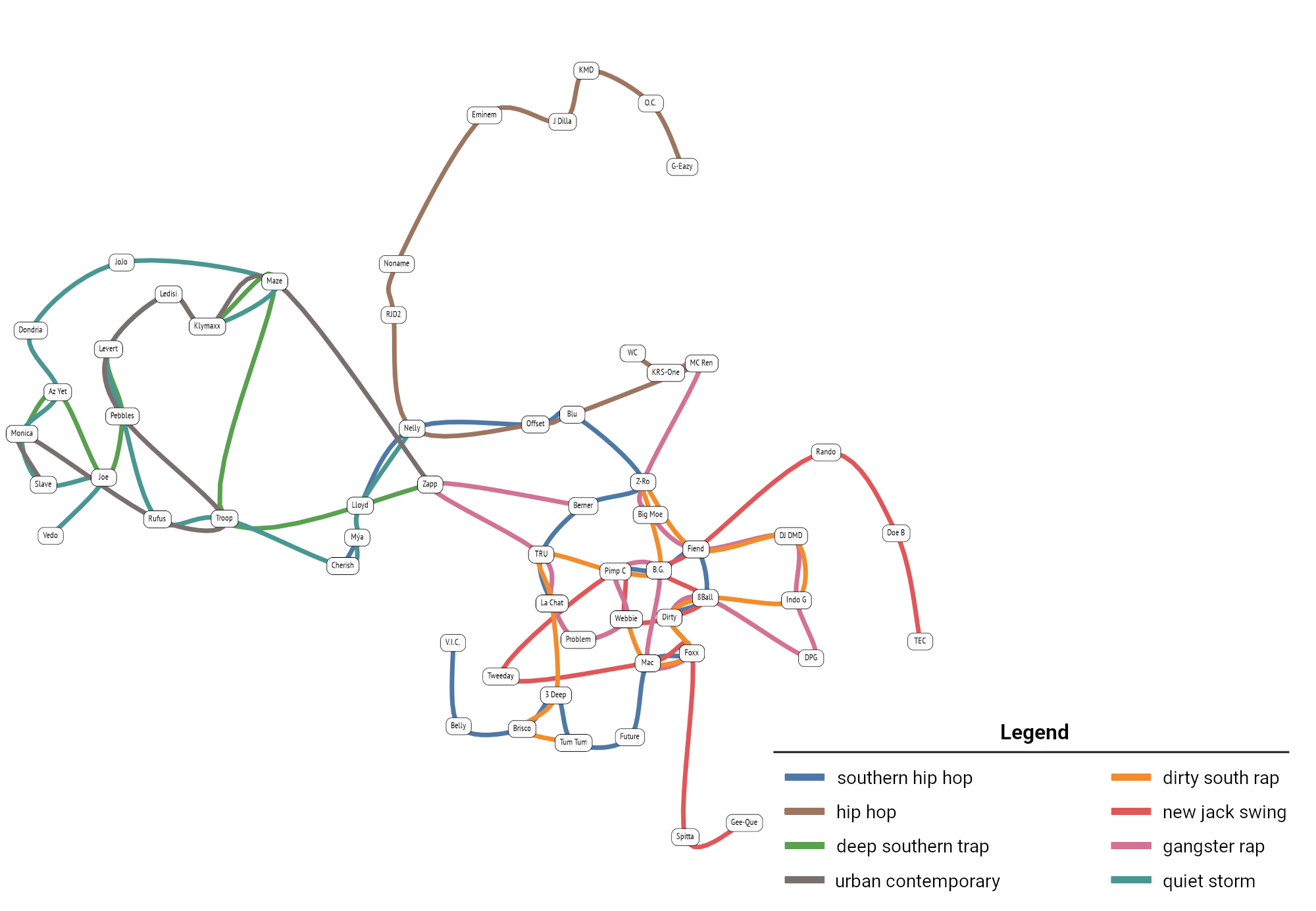}
	}
		\subfloat {
		\includegraphics[width=0.33\textwidth]{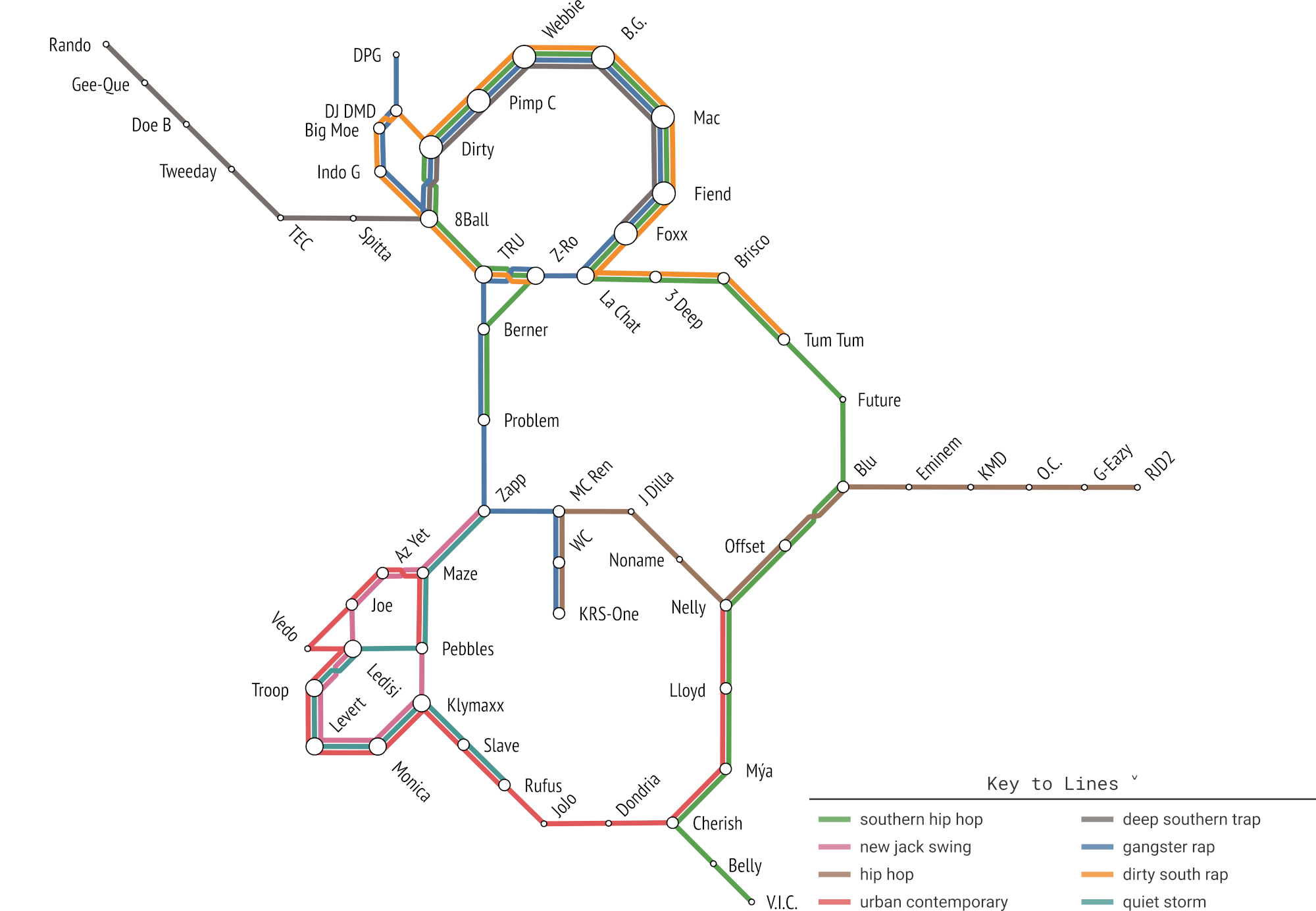}
	}
    \caption{Our study evaluates the readability of set visualizations for abstract, non-spatial data generated by publicly available visualization systems (EulerView (left), LineSets (middle), MetroSets (right)) by measuring their performance on representative element-based and set-based tasks. }
    \label{fig:teaser}
\end{figure*}

With this in mind, our search narrowed down to three set visualization systems: EulerView~\cite{saa-favos-09}, a technique representing the popular class of Euler diagrams, LineSets~\cite{ahrc-dslnvt-11}, a technique representing overlay-based set visualizations, and MetroSets~\cite{jwkn-mvsmm-20}, a graph-based technique using the metro map metaphor; see Figure~\ref{fig:teaser}. 

For our evaluation, we performed an online experiment with a total of 120 participants who solved six different tasks on stimuli of two sizes generated by the three selected systems. 
We used
small ($|\mathcal U|=30$ and $|\mathcal E| = 6$) and large ($|\mathcal U|=60$ and $|\mathcal E| = 8$) 
real-world set system data derived from Spotify's database of artists and genres.

Note that the terms ``small" and ``large" are used here simply to distinguish between the two sizes, rather than as descriptive of the actual sizes of the underlying data. Nevertheless, such small datasets do occur in research papers, popular science, and social media, whereas truly large datasets require visual analytic systems.

The  results show statistically significant quantitive and qualitative differences, suggesting that MetroSets scales better and performs more consistently than EulerView or LineSets, and is also better-liked by users. 







\section{Related Work}

\subsection{Set Visualization Techniques}
The survey of Alsallakh et al.~\cite{alsallakh2014visualizing} provides a classification of set visualization techniques into six different groups.
The most commonly used techniques belong to the family of \textbf{Euler and Venn diagrams}, which represent each set as a region enclosed by a simple, closed curve and set relations by corresponding intersections and inclusions of these regions. 
Most of these diagrams focus on representing the sets, but not necessarily the individual elements. 
Only two Euler-based techniques show individual elements~\cite{saa-favos-09,rd-ued-10}.
Area-proportional Euler diagrams~\cite{larsson2020eulerr,w-eaacved-12,DBLP:conf/diagrams/RodgersFSH10} do not show the elements, but indicate set cardinalities by scaling the region sizes proportionally.
Some methods for generating Euler diagrams use simple shapes such as circles and ellipses to represent the sets~\cite{sfrh-adedwc-12,w-eaacved-12,larsson2020eulerr,mr-efled-14}, others are more widely applicable but also produce more complex shapes~\cite{rd-ued-10,saa-favos-09,sas-sabied-16,rzf-gedg-08,srhz-iged-11}.
Well-formed Euler diagrams do not always exist and some techniques may produce inconsistent diagrams showing non-existing set intersections.

The second class of techniques are \textbf{overlay-based set visualizations}. 
Here, all the elements are explicitly shown at pre-defined positions in the plane, which could represent geolocations in spatial datasets or positions of nodes in a node-link diagram of underlying network data. 
The techniques create visual overlays that indicate the set memberships of the input point set. 
Bubble Sets~\cite{cpc-bsrrwioev-09} 
create isocontours for each set, which enclose their corresponding points similarly to regions in Euler diagrams.
Pairs of isocontours, however, may intersect even if their two sets have an empty intersection in the data. 
LineSets~\cite{ahrc-dslnvt-11} is based on the idea of connecting the elements of each set by a smooth and short curve reminiscent of lines in metro maps.
Kelp Diagrams~\cite{dksw-kdpmv-12} and KelpFusion~\cite{mhsad-khvt-13} extend and combine ideas from LineSets and Bubble Sets by parameterizing set representations from sparse spanning graphs to convex hulls, with the middle range resulting in bubble shapes for local point clusters in a set and edge-like links to connect elements further apart.
All overlay techniques inherently require a given embedding of the set elements and so these methods are not directly applicable for abstract set systems. 

\textbf{Visualizations based on bipartite graphs} represent elements and sets as two types of vertices.
Each element is connected by an edge to all sets containing that element, which yields a bipartite graph. 
General visualization techniques for (bipartite) graphs can then be employed and have been integrated in several systems~\cite{sgl-jsiativ-08,drrd-pstfis-12,be-dhss-00,aamh-rsivalos-13}.
However, for complex set systems, the resulting layouts are dense with many edge crossings and there is little support for set-related tasks~\cite{alsallakh2014visualizing}.

Due to their equivalence to set systems, \textbf{hypergraph layout methods} are also relevant. 
Johnson and Pollak~\cite{jp-hpcdvd-87} introduced different notions of hypergraph planarity. Many papers study, from a theoretical perspective, support graphs (or supports in short), which are graphs defined on the vertex set $\mathcal U$, such that the elements of each hyperedge $S\in \mathcal E$ induce a connected subgraph. A drawing of the support graph could then be used as the basis for visualizing the set system. 
Of particular interest are planar supports~\cite{bkmsv-psh-11a,cgmny-spssh-19}, tree supports~\cite{ks-cmoct-03,kmn-mtshled-14}, and path supports~\cite{bcps-psh-12}.
Not all hypergraphs admit all types of (planar) supports, which may limit their applicability in practice.
The recent MetroSets system~\cite{jwkn-mvsmm-20} computes a (not necessarily planar) path support as the basis for rendering abstract set systems in the style of a metro map.

Another class of techniques are \textbf{matrix-based approaches}, which represent sets and elements as the rows and columns of a matrix.
An element contained in a set is indicated by coloring or marking the respective matrix cell with a glyph.
More complex set-related tasks typically require interaction, and most matrix-based systems are designed for interactive analysis and exploration.
Examples are ConSet~\cite{kls-vcwpmd-07}, OnSet~\cite{smds-ovtlbd-14}, RainBio~\cite{lamy2019rainbio}, or UpSet~\cite{lex2014upset}.
Matrix-based methods scale well, but have a strong dependency on row and column ordering, are less intuitive for non-experts, and require interaction with a non-trivial interface for the more complex set visualization tasks.
Closely related are \textbf{linear diagrams}~\cite{rsc-vswld-15,lm-clmdv-19,stapleton2019efficacy}, where again each set is a row in a table. 
The columns, however, do not represent individual elements, but rather non-empty set intersections. 
Any marked or colored cell in the table indicates that a set is part of a non-empty set intersection with all the other sets being marked in the same column.


Finally, there are many \textbf{aggregation-based techniques} implemented as components in interactive visual analytics applications, e.g.~\cite{aamh-rsivalos-13,  DBLP:journals/tvcg/FreilerMH08,   DBLP:journals/tvcg/KosaraBH06}.
These techniques are designed for analyzing large-scale data and aggregate elements into groups, while maintaining all set relations. Since they do not support element-based tasks, we did not consider them for our study.



\subsection{Evaluations of Set Visualization Techniques}

Several aspects of set visualizations have been empirically studied in the literature before, but are  limited to evaluating parameters of a single visualization style, comparing overlay-based techniques with pre-embedded elements only, or excluding element-based tasks completely.
All studies mentioned below build their experimental design on static images generated with their respective target systems and approaches (like ours).
Alper et al.~\cite{ahrc-dslnvt-11} performed a readability evaluation of their overlay-based system LineSets in comparison to Bubble Sets~\cite{cpc-bsrrwioev-09}. 
In their study 12 participants performed four element-based and set-based tasks on spatial and social network data with 3--5 sets and 50--200 elements; accuracy and time were recorded. 
For detailed tasks and preference ratings, LineSets outperformed Bubble Sets; otherwise there were no significant differences.
Meulemans et al.~\cite{mhsad-khvt-13}  evaluated KelpFusion, LineSets, and Bubble Sets with 13 participants. Their geographic datasets had  4 or 5 sets and 12--49 elements; the four tasks were similar to those 
by Alper et al.~\cite{ahrc-dslnvt-11}. The results showed no significant difference in performance or preference between KelpFusion and LineSets, but both outperformed Bubble Sets.
Both studies considered only pre-embedded set systems.


Two studies evaluated techniques that show combined set and network visualizations. Rodgers et al.~\cite{DBLP:journals/isci/RodgersSAMBT16} included five different systems in their crowdsourced study: Bubble Sets, LineSets, KelpFusion, EulerView, and their  system, SetNet. They used datasets with 11--64 elements, 3--7 sets, and 42--162 edges in the network. All tasks involved both the network structure and the set information. 
Their results indicated that SetNet and EulerView significantly outperform the other three systems on the combined set+network tasks.
Baimagambetov et al.~\cite{DBLP:conf/diagrams/BaimagambetovS020} compared SetNet, Bubble Sets, and WebCola\footnote{see \url{https://ialab.it.monash.edu/webcola/}} on instances with 2--8 sets, 10--100 elements, and 40--170 network edges. Their study did not assess readability, but rather quantitatively evaluated the frequency of inaccuracies and properties that have previously been empirically confirmed to be visually ineffective. Bubble Sets turned out to be least inaccurate system and in terms of ineffectiveness there was no clear winner.

Further evaluations considered techniques such as Euler diagrams, linear diagrams, and mosaic diagrams, all of which do not represent individual elements and thus used only set-based tasks~\cite{csrmb-vsecdt-14,lm-clmdv-19}. They showed that linear diagrams outperformed Euler diagrams and were on par with mosaic diagrams.


Thus, despite several previous specialized evaluations, a broader study of abstract set visualization across different classes of techniques as well as including both set- and element-based tasks is missing.
Hence the observation of Alsallakh et al.~\cite{alsallakh2014visualizing} that there is a lack of empirical studies assessing the effectiveness of set visualization techniques, remains valid, especially regarding fundamental element-based and set-based tasks on abstract set data using visualizations that show the individual set elements.






\section{Principles of Abstract Set Visualization}\label{sec:principles}

The design of set system visualizations is guided by the types of tasks they should support, as well as by the type of information that needs to be represented. 
Alsallakh et al.~\cite{alsallakh2014visualizing} present a taxonomy of tasks classified into the following three categories.

\begin{compactenum}
    \item \textbf{Element-based tasks} are concerned with specific elements and their respective relationship to the sets. For example: What music genre(s) does 'Van Halen' belong to?
    \item \textbf{Set-based tasks} are concerned with the relationship between different sets without taking individual elements into account. For example: Which music genres overlap with 'Rock'?
    \item \textbf{Attribute-based tasks} are concerned with attributes of set elements and their relationship of distribution in regards to set membership. For example: Do artists in the 'Rock' genre sell more records than artists in the 'Hip Hop' genre? 
\end{compactenum}

We focus on element-based tasks and set-based tasks in this study as they represent elementary tasks on abstract set systems that can be performed on static (rather than interactive) visualizations that appear in research papers, newspaper articles and social media. 
According to~\cite{alsallakh2014visualizing} the support of the different types of tasks is closely tied to the type of information represented in the visualization.

\begin{compactitem}
    \item \textbf{Representing set information only:} the focus is on the relationships between sets and individual elements might not be explicitly represented.
    \item \textbf{Representing individual elements:} Individual elements are represented explicitly, making it possible to also encode additional attribute information.
\end{compactitem}

One important aspect of set visualization techniques is their visual scalability, i.e., whether a visualization remains comprehensible when the number of elements and/or the number of sets increases.
Naturally, scalability is closely related to the size of the set system.
However, scalability also depends on the information that is represented. For example, visualizations that only represent set information are not adversely affected by increasing the number of elements.
Even when only set information is represented, scalability depends not only on the visual metaphor, but in some cases (e.g., Euler and Venn diagrams) also on the structure and number of the intersection relationships of the underlying set system.

Explicitly representing elements is associated with increased visual complexity as individual elements need to be depicted.
Furthermore, it is non-trivial to assign primacy to either the depiction of the sets or of the elements, as one can influence the comprehension of the other negatively.

\section{Systems}\label{sec:systems}

In our study we focus on systems that produce static visualizations that can be used to argue about the set system and to perform elementary tasks, rather than interactive visual analytics systems.
The general requirements determining whether a given system could be used in our study include the following criteria:

\begin{compactenum}
    \item All elements are explicitly depicted.
    \item All elements are identifiable by an associated label.
    \item All sets are identifiable by a given label or legend.
    \item No interactivity is required to reason about relationships between sets and between elements and sets.
    \item The visualization system is implemented and its source code available.
\end{compactenum}

Most systems that were taken into consideration do not fulfil all of the above criteria: some do not depict all elements (e.g., Eulerr~\cite{larsson2020eulerr}, VennEuler~\cite{w-eaacved-12}, Linear Diagrams~\cite{rsc-vswld-15}), others do not label the elements (e.g., eulerGlyphs~\cite{DBLP:journals/tvcg/MicallefDF12}), still others focus on different tasks (e.g., network based tasks such as SetNet~\cite{DBLP:journals/isci/RodgersSAMBT16}). Many do not provide implementation (e.g., untangling Euler diagrams~\cite{rd-ued-10}). 
Eventually we narrowed our selection down to three systems: EulerView~\cite{saa-favos-09}, LineSets~\cite{ahrc-dslnvt-11}, and MetroSets~\cite{jwkn-mvsmm-20}.

\textbf{EulerView} is an Euler-based technique. Notably, while many Euler diagram systems do not represent individual elements, EulerView explicitly visualizes elements and their respective labels. 

\textbf{LineSets} is a representative of the overlay-based visualization approaches. As such, it requires initial positions for the elements.
To resolve this issue, we contacted the authors and followed their suggestion to pre-process the data with a force-based layout~\cite{fruchterman1991graph}.

\textbf{MetroSets} is visually similar to LineSets but is designed with abstract data in mind. It is the most recent of the three techniques and combines ideas from  graph-based and overlay-based approaches.

While other overlay-based systems also meet our criteria (e.g. BubbleSets~\cite{cpc-bsrrwioev-09}), we chose to use LineSets because of its strong conceptual and visual similarity to MetroSets.

For more detailed descriptions and implementation details of the three selected systems we refer the reader to the supplemental material.

\section{Controlled Study}\label{sec:study}

As mentioned in~\autoref{sec:principles} representing information is closely coupled with tasks that can be performed on a set visualization and therefore task  performance can give a general idea about the effectiveness of the visualization. Traditionally, task performance is measured by accuracy (number of tasks solved correctly) and task completion time, with better performance associated with high accuracy and/or low completion time. 
The aim of our study is to evaluate the effectiveness
of the three visualization systems 
on a set of carefully selected tasks that span the spectrum of elementary element-based and set-based tasks.
We assessed the different systems by performing a controlled human subjects study with two experiments (that differ in the size of the underlying data). We collected quantitative data (time and accuracy) as well as qualitative data (Likert-scale subjective ratings).

While lab-experiments can be better controlled, we opted for a fully online setting due to difficulties associated with in-person gatherings in the year 2020. With this in mind, we used social media posts (e.g., reddit, facebook, instagram) to recruit participants. We stopped collecting data after reaching 120 fully completed instances. 
A repository with all our datasets, stimuli, responses of participants and evaluation code can be found on OSF\footnote{see \url{https://osf.io/nvd8e/}}.

\subsection{Participants and Setting}

\begin{figure}
  \centering
  \subfloat {
    \includegraphics[width=0.49\linewidth]{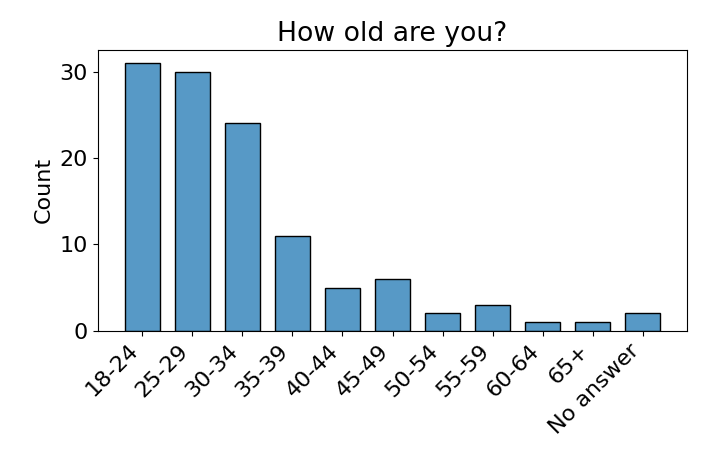}
  }
  \subfloat {
    \includegraphics[width=0.49\linewidth]{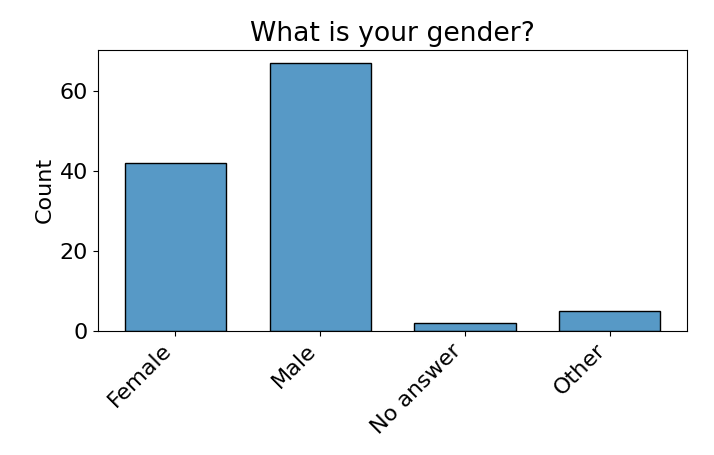}
  }
  \qquad
  \subfloat {
    \includegraphics[width=0.49\linewidth]{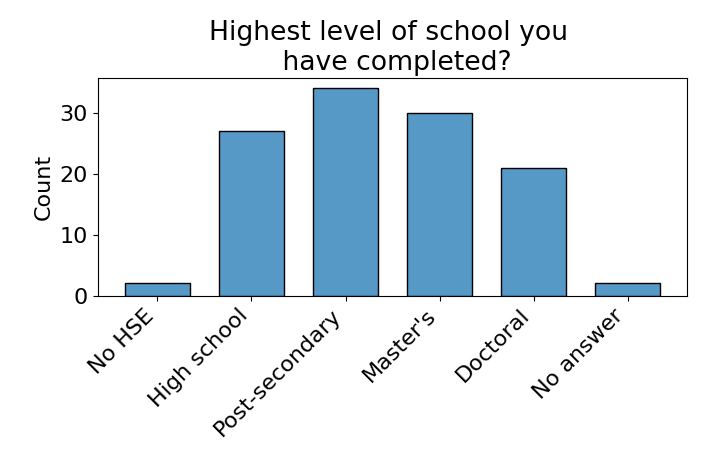}
  }
  \subfloat {
    \includegraphics[width=0.49\linewidth]{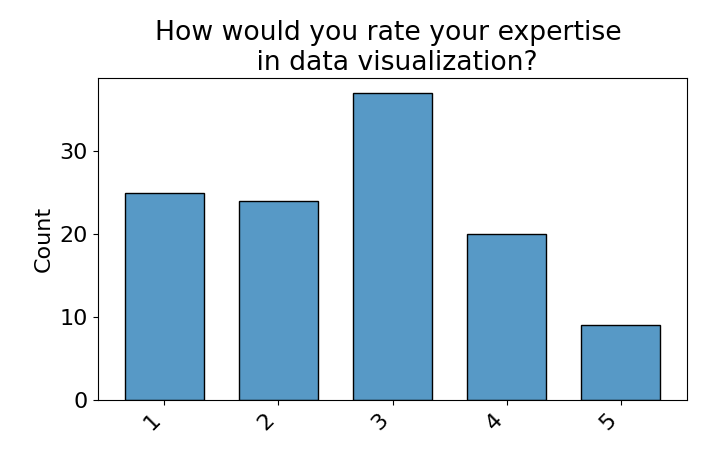}
  }
  \qquad
  \subfloat {
    \includegraphics[width=0.49\linewidth]{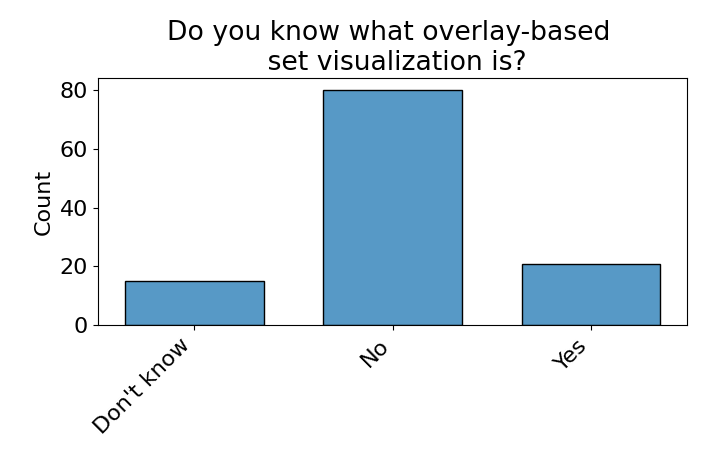}
  }
  \subfloat {
    \includegraphics[width=0.49\linewidth]{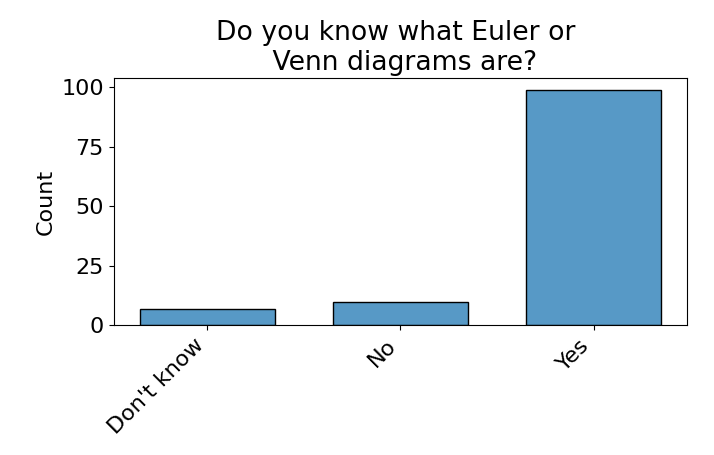}
  }
  \caption{Demography of the 120 participants.\label{fig:demo}}
\end{figure}

Participants in the study were at least 18 years of age, without known color vision deficiency. 
53\% of participants reported being younger than 30, although we had non-trivial numbers of participants aged 55 or older. 
57\% of the participants were male, 36\% were female, and the rest identified as other or preferred not to answer. 
The majority of the participants were well-educated (73\% with a post-secondary degree or higher) but did not report high familarity with data visualization systems (average self-rating of 2.7 out of 5.0). 
Nevertheless, many were familiar with Euler or Venn diagrams, and some with overlay-based set visualization.
A summary of the demographic data is in~\autoref{fig:demo}.

The study itself was realized using the LimeSurvey open-source online survey tool;\footnote{see \url{https://www.limesurvey.org/en}} and hosted online. 
We used the standard functionality of question types provided by LimeSurvey but modified the layout to place images and questions side-by-side to reduce the need for scrolling; see~\autoref{fig:t5euler}. 

In total 44 question, excluding transition screens, had to be answered to complete the study: 7 preliminary screening questions, 8 demographic questions, 6 task training questions, 18 testing questions, and 5 qualitative questions.

The participants were randomly assigned to either the small or large experiment (details about the two are provided in Section~\ref{sec:sizes}).


\subsection{Datasets}\label{sec:dataset}
We used a different real-world dataset for each task to minimize the variance that might result from one dataset being well-suited for a certain system.
We created a hypergraph (4549 hyperedges, 347686 vertices) of artists and music genres of Spotify data and automatically extracted subgraphs which where subsequently used to generate the stimuli in all three systems.
In the context of the Spotify hypergraph, artists are set elements represented by vertices, and genres are sets represented by hyperedges. 

We used the following approach to extract the subgraphs. We set a target number of elements and sets; 30 elements and 6 sets for the small datasets and 60 elements and 8 sets for the large dataset. 
We also specified the number of sets that elements should belong to. 
For subgraphs of the small dataset we required the graph to have ten elements that are members in exactly one set, ten elements that are members in two sets, six elements that are members in three sets and four elements that are members in four sets. 
For the large dataset we doubled the number of elements in each membership degree category, but did not add another category.

To extract a single subgraph we manually declared a seed set in the hypergraph and iteratively added more sets that overlap with the current sets until the target number of sets was reached, allowing for enough elements to be added in each membership category. 
We proceeded by iteratively picking random elements from the set of six or eight sets that fulfilled the membership requirement until the target number of vertices was reached. If no valid subgraph was extracted after a fixed number of iterations, we concluded that none existed and started the process again with another seed node.

The previous step gave us a potential candidate that had to fulfill further requirements. 
First, we excluded hypergraphs with multiple disjoint components.
Second, to avoid trivial instances, we excluded hypergraphs with fewer than three elements per set.
Third, as the study targeted an English-speaking audience, we also excluded hypergraphs with vertices containing non-Latin characters in their labels.

After generating the first set of stimuli we decided to additionally exclude elements with long labels (greater than eight for the large dataset and greater than twelve for the small dataset). 
The reasons behind this decision are that all three systems struggle handling long labels, and long individual labels distract from the overall visualization. 

With our extraction approach we were able to create pseudo-realistic datasets from real-world data with the advantage of having control over relevant properties. Note that it would have been much easier to use a synthetic dataset. We considered that possibility, but decided against it for several reasons. Primarily, we believed that using real-world data would make the study more enjoyable, thereby increasing  
participation. We discuss potential problems with this approach in~\autoref{sec:limitations}.

\subsection{Size}
\label{sec:sizes}
We determined the two dataset sizes (small and large) based on observations we made in the three systems during the pilot experiment. Our goal was to choose sizes which reflected the capabilities of all three systems, while avoiding situations where the extreme difficulty (or ease) of a task might obscure the differences between systems.

Recall that the terms ``small" and ``large" are used here simply to distinguish between the two sizes, rather than as descriptive of the actual sizes of the underlying data. The size of the small dataset (30 elements and 6 sets) can be considered a typical size for set systems where naive approaches, e.g. classic Euler diagrams, start to struggle. We determined the size of the large dataset (60 elements and 8 sets) to be the largest such that the resulting visualizations were consistently readable on an ordinary monitor without user interaction.

\subsection{Tasks}
We used the taxonomy of Alsallakh et al.~\cite{alsallakh2014visualizing} to select elementary tasks of set systems that participants would perform.
Specifically, we selected three element-based (T1--T3) and three set-based (T4--T6) tasks 
that can be performed on all three systems without the need for interaction.
We re-worded technical terms (e.g., element, set, hypergraph) using a more natural language for better accessibility.

\begin{compactitem}
    \item[\textbf{T1}] \textbf{Find/Select elements that belong to a specific set}: Check all of the 'Genre' artists below; Three artists were given as possible answers.
    \item[\textbf{T2}] \textbf{Find sets containing a specific element}: What genre(s) does 'Artist' belong to; All sets were given as possible answers.
    \item[\textbf{T3}] \textbf{Find/Select elements based on their set memberships}: Check all of the artists below that belong to both 'Genre 1' and 'Genre 2'; Three artists were given as possible answers.
    \item[\textbf{T4}] \textbf{Analyze intersection relation}: Please check below if any artist(s) belong to both of the following pairs of genres; Three pairs of genres were given as possible answer.
    \item[\textbf{T5}] \textbf{Identify set intersections belonging to a specific set}: Which genres overlap with 'Genre'; All sets except the specified one were given as possible answers.
    \item[\textbf{T6}] \textbf{Analyze and compare set- and intersection cardinalities}: How many artists are both 'Genre 1' and 'Genre 2'; Numbers from 0--10 were given as possible answers.
\end{compactitem}

We assigned one small and one large dataset to each task and used different approaches to select a balanced set of possible answers. We excluded tasks which were ambiguous or impossible to solve (e.g., nodes overlapping lines in MetroSets, overlapping labels in EulerView, covered up lines between two elements in LineSets).

As our question types were limited to those provided by LimeSurvey, we elected to use multiple choice for every task except for T6.
Each option in a multiple choice question can be thought of as a subtask and we counted tasks T1-T5 as correct only when a participant answered all subtasks correctly.

Task T2 required a participant to select all sets an element is member of. For the sake of fairness we picked elements with the exact same set membership degree.
The same reasoning was applied to task T5, where participants had to select all sets intersecting with a given set, and we selected sets that had the same cardinality of intersections.

Lastly, participants had to correctly count the number of elements in the intersection of two sets in task T6. We selected pairs of sets with a similar number of elements (6-8) in their intersections.

\subsection{Stimuli}

We generated static images, or stimuli, for all combinations of dataset and system. The images for MetroSets and EulerView were rendered in their native implementation. The `balanced' pipeline preset was used for MetroSets. For LineSets, our implementation provided output in the form of a dot file. The final visualization was then rendered to a PNG file using graphviz\cite{ellson2001graphviz}, with parameters chosen to match those used in the original paper introducing LineSets\cite{ahrc-dslnvt-11}. To minimize possible confounding factors, we used the color scheme from MetroSets for our LineSets stimuli.


We modified all three systems to use the same font: \textit{PT Narrow Sans}. The EulerView module automatically removes label overlaps by only showing a subset of non-overlapping labels.
As this approach is not suitable for our study, we fixed the font size to the largest possible so that we could remove overlaps by setting the label anchor to the left, right, or top of the glyph (instead of only the bottom by default). 

As all systems generate slightly different output, we used image manipulation software to normalize the difference. 
We created a blank image with 2000px width and 1385px height. We then scaled the images generated by the systems to fit into the canvas with space to fit a legend into the bottom right corner of the image. In the case of LineSets, we found that it was often impossible to fit the stimuli alongside the legend without making labels unreadably small. To resolve this, we automatically rotated the initial layout provided to LineSets by 90 degrees whenever doing so reduced its height. We did not encounter this issue with the other systems, and therefore did not rotate any other stimuli.

MetroSets is the only system that automatically creates an image that has a legend.
While sets in~\cite{saa-favos-09} are by default annotated with a label, we decided to remove any potential confounding factors and instead created a legend with two columns of circular (50px diameter) texture samples and the associated label.
We created a similar legend for LineSets using the same design as MetroSets legend.
All legends use 22pt \textit{Roboto Mono} as header and 18pt \textit{Roboto} for set labels.


To reduce the time required to find elements in the visualization for tasks T1-T3, we highlight the relevant elements by adding a colored 2px rectangular border around them; see~\autoref{fig:t5euler}. The main reason for adding highlighting was to exclude the time required to
search for elements, as this  could be considered its own task and would add noise to our overall time measurements.

We applied an additional sharpening effect to all EulerView stimuli, as the images tended to be blurry after normalization. 


We communicated with the authors of all three systems to ensure that we generate visual stimuli that represent each system fairly. We also discussed the slight modifications we made (e.g., using the same fonts, using the same colors, reducing label overlaps). 
All visual stimuli used in the study can be found in the supplemental material.

\subsection{Experimental Procedure}

For each condition (EulerView, LineSets, MetroSets) we conducted two experiments, one on the small and one on the large dataset. 
The experimental section of the study had three set-based tasks and three element-based tasks. This resulted in $3 \: \mathrm{conditions} \times (3 + 3) \: \mathrm{tasks} = 18 \: \mathrm{trials}$. Participants were randomly assigned to work with either small or large datasets.

Both experiments follow the five-phase template; (1) information consent and screening, (2) demographic questions, (3) tutorial, (4) formal study and (5) post-task questionnaire. 

In phase one participants were given the background information and procedure of the study before they had to give consent to the study policy. 
The policy stated that only participants of age 18 and above with no known color vision deficiency are allowed to participate. 
After giving consent the participants were first shown an image with excerpts of all three visualization styles. 
The excerpts show the worst case label size and we ask a yes/no question if the labels are readable in all three systems. 
Additionally, we state that the window size should not be altered during the study and automatically logged the window information. 
We assume that participants did not change their screen size during the experiment. 
This information was used to screen and exclude participants who were potentially not able to accurately identify set elements.
Afterwards the participant completed a subset of six plates of the Ishihara Test~\cite{clark1924ishihara} to screen for potential issues with color perception. 
Failing to give the correct answer could be due to different potential causes (e.g. viewing angle or uncalibrated monitor).

In phase two 
the participants provided (optional) demographic information, summarized in~\autoref{fig:demo}. 
In phase three the participants were introduced to the different tasks and visual stimuli.
We had a total of six tasks: three set-based tasks and three element-based tasks. 
We randomly assigned a set-based task and a element-based task to each system.
Participants were not able to proceed to the next task until a correct answer was given.
The tutorial portion of the study does not contribute to the results.

\begin{figure}
    \centering
    \includegraphics[width=\columnwidth]{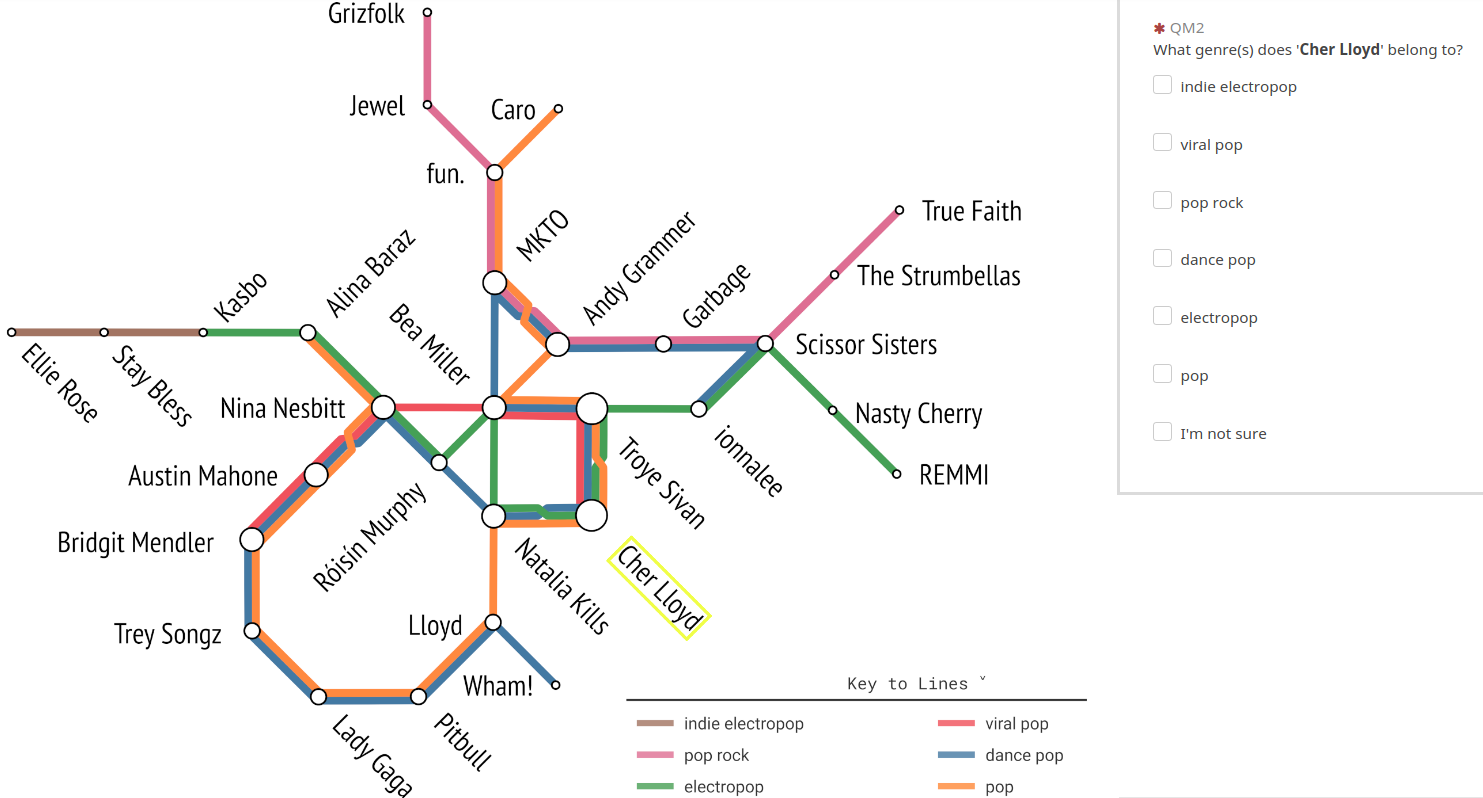}
    \caption{Example of a element-based task (T2) in combination with the MetroSets system.}
    \label{fig:t5euler}
\end{figure}

Phase four contains the quantitative portion of the study. Each participant was presented with 18 tasks split into 6 blocks of three tasks.
Each block consisted of exactly one question per system of different types of our potential set of 6 tasks; for an example task see~\autoref{fig:t5euler}.
We randomized the order in which the blocks were shown to the participant, as well as the order of questions inside each block.
Before each block the participants were shown a white screen with text letting them know that they can take a break.

In phase five the participants provided qualitative feedback with four Likert scale questions and one fillable text field. Specifically,  we asked the following questions:

\begin{compactitem}
    \item How clearly could you identify to which genre(s) an artist belonged to?
    \item How clearly could you identify the overlap between different genres?
    \item How often did you use pre-existing knowledge about music to answer an question? 
    \item How interested are you in using each style again? 
    \item Please share any thoughts you have about the different styles!
\end{compactitem}



\subsection{Pilot Study}
Before recruiting participants we asked a group of four experts and six lay people to give their opinion on the design of the study. We asked the lay people for their opinions on duration, wording, and experience.
The experts where given a questionnaire and asked for further comments/suggestions.
From the information gathered we made several modifications to the study design.

The first major change was to not include all tasks of the small and large dataset in a single experiment, as the study duration was too long (about 45 minutes on average) and caused participants to tire.
As a result, we split the experiment into two experiments: one with the small dataset and another with the large dataset.
We also reduced the number of questions we asked for each task.

As suggested by several of the experts, we added highlighting for all element-based tasks (T1, T2, and T3). Otherwise the time required to search for the relevant elements dominates the time measurements.

Initially, T4 asked a participant to identify if a pair of given sets intersect with a simple yes/no question. 
We modified this question, as it had a $50\%$ chance of being guessed correctly, and instead asked if three presented pairs of genres intersect. 

The qualitative evaluation was streamlined and shortened from eight Likert type questions to four.
We also added a free form text field,
asking the participants to leave comments.

\subsection{Hypotheses}\label{sec:hypotheses}

The goal of our study is to evaluate the differences in readability in terms of the performance of both element-based and set-based tasks among abstract set visualization techniques. We selected one representative each from the three classes of techniques~\cite{alsallakh2014visualizing} that show individual elements and are suitable as visualizations for non-experts, namely Euler diagrams, overlay-based techniques, and graph-based techniques.

We formulate three different hypotheses regarding element-based task performance, set-based task performance, and visual scalability. For the purpose of this study, we say that a technique $A$ outperforms a technique $B$ for some task(s) of interest if the results of $A$ are significantly more accurate than $B$, using a significance threshold of $0.05$.
Further, we say that technique $A$ scales better than $B$ if the accuracy of $B$ decreases significantly more than the accuracy of $A$ when used with a larger dataset.

MetroSets and LineSets use a similar visualization style, but MetroSets can achieve lower visual complexity by freely placing the elements and schematizing the line trajectories. In addition, previous studies have shown advantages of LineSets over Euler-like visualizations~\cite{mhsad-khvt-13,ahrc-dslnvt-11}. Finally, as we observed in \autoref{sec:systems}, both LineSets and EulerView can sometimes produce ambiguous results, and we expect this to occur more frequently with more complicated datasets. Based on these facts, our hypotheses are as follows:


\begin{itemize}
    \item \textbf{H1}: For element-based tasks MetroSets will outperform LineSets and LineSets will outperform EulerView.
    \item \textbf{H2}: For set-based tasks MetroSets will outperform LineSets and LineSets will outperform EulerView.
    \item \textbf{H3}: MetroSets will scale better than LineSets and EulerView.
\end{itemize}

\subsection{Summary of Supplementary Material}

We provide all related material of the study in the supplementary materials: 
the code for dataset extraction and input format parsing, stimuli for pilot and study, images of a full walkthrough of the study, the raw results, the statistical analysis code, and the results of our ANOVA tests. The code for MetroSets can be found on \href{https://osf.io/nvd8e/}{OSF}, the LineSets implementation in GMAP is available on \href{https://github.com/spupyrev/gmap}{GitHub}. We used \href{https://tulip.labri.fr/TulipDrupal/}{Tulip 3.5} to build and run the EulerView module.

\section{Results and Analysis}

\subsection{Data}

After filtering out participants who failed to complete the entire study or gave incorrect answers on the Ishihara Test, we were left with a total of 116 respondents. Exactly half of these respondents worked with small datasets, while the other half performed tasks with large datasets.



\subsection{Methods}

To analyze the relationship between style of visualization, dataset size, and accuracy, we performed a 2-way ANOVA test \cite{fisher1992statistical}, including an interaction term meant to capture the differences in scalability between the systems. For each statistically significant result ($\alpha = .05$), we performed post-hoc analysis using Tukey's test \cite{tukey1949comparing}. This experiment was repeated separately for each of the 6 tasks we asked participants to perform.

\subsection{Accuracy}

\begin{figure*}
    \centering
	\subfloat {
		\includegraphics[width=0.33\textwidth]{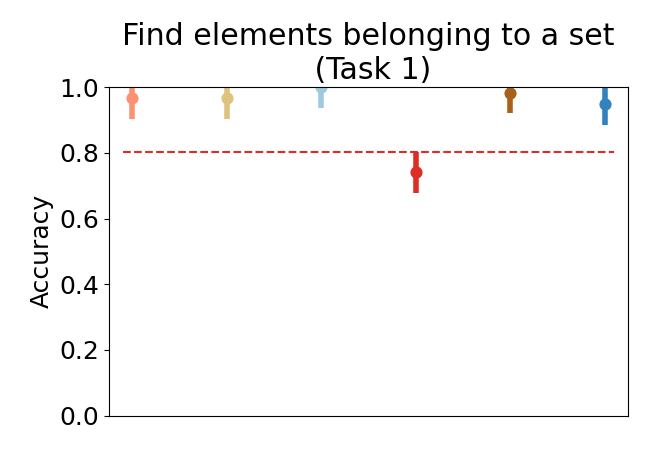}
	}
	\subfloat {
		\includegraphics[width=0.33\textwidth]{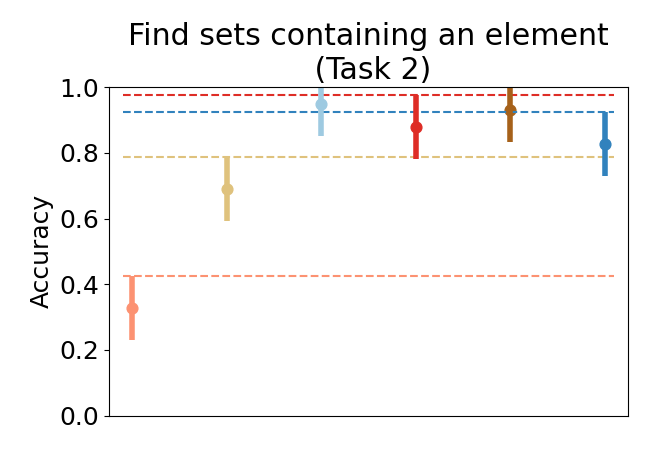}
	} 
	\subfloat {
		\includegraphics[width=0.33\textwidth]{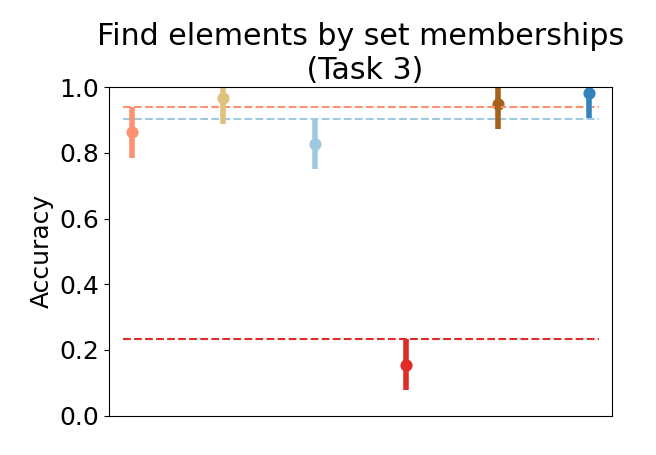}
	} \\
	\subfloat {
		\includegraphics[width=0.33\textwidth]{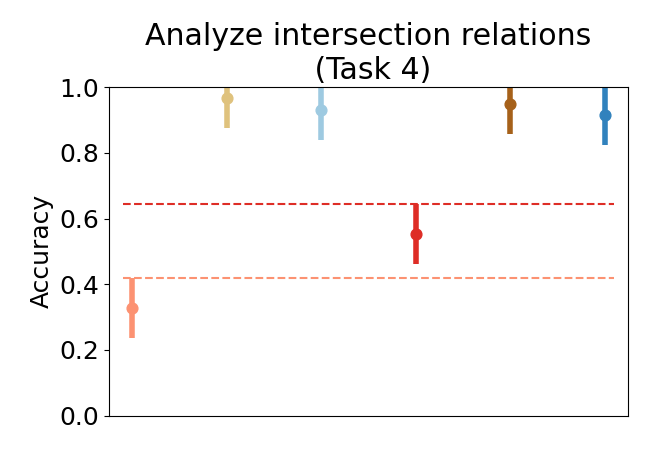}
	} 
	\subfloat {
		\includegraphics[width=0.33\textwidth]{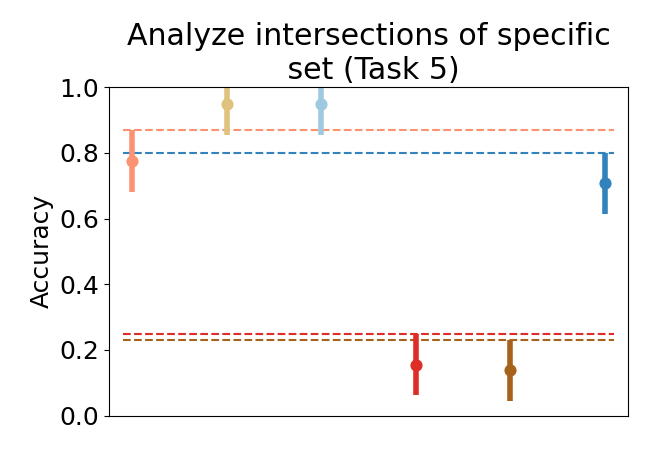}
	}
	\subfloat {
		\includegraphics[width=0.33\textwidth]{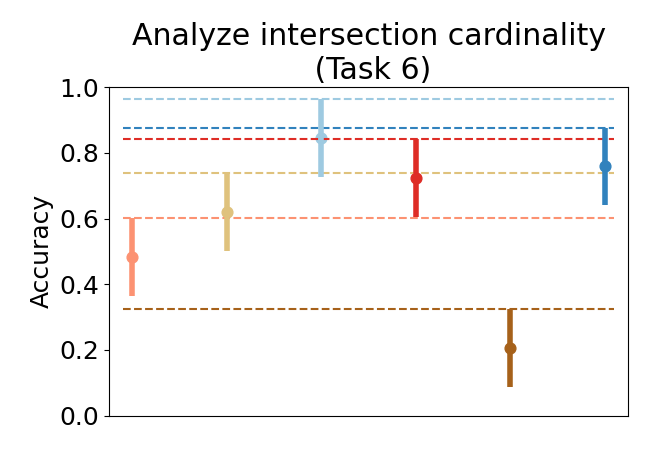}
	} \\
	\subfloat{
		\includegraphics[width=0.5\textwidth]{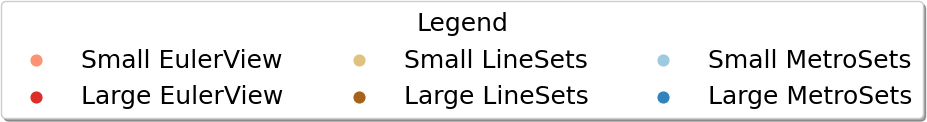}
	}
	\caption{Task accuracy results: each plot corresponds to a single task, and contains means and adjusted 95\% confidence intervals for the accuracy associated with each pair (size, system). A horizontal guideline has been added to the top of each confidence interval to make it easier to determine if two intervals overlap; if the interval for one system lies completely above the guideline for another system, then it means the first system performed significantly better ($\alpha = 0.05$). Likewise, if the guideline for one system passes through the interval for another system, then the two are not significantly different.}
	\label{fig:tukey-results}
\end{figure*}

Our ANOVA showed that all factors were statistically significant for all tasks, with the exception of size, which did not have a significant impact for task T4 or T6. Post-hoc testing confirmed significant differences on all six tasks, which are presented in \autoref{fig:tukey-results} and described explicitly below.

On task 1, there was no significant difference between the three systems on the small dataset. However, on the large dataset, EulerView performed significantly worse than either MetroSets or LineSets ($p < .001$, mean difference $\approx 22\%$).

Conversely, on task 2, there was no significant difference between the three systems on the large dataset. However, on the small dataset, MetroSets was significantly more accurate than LineSets ($p=.002$, mean difference $\approx 25\%$), which in turn was significantly more accurate than EulerView ($p<.001$, mean difference $\approx 36\%$).

On task 2, the participants performed better with the larger dataset. Looking more carefully at the stimuli provides a plausible explanation. The smaller dataset here contained four overlapping sets to be identified,
while the larger dataset contained only three. This is likely the cause of the difference observed: a priori, increasing the size of the visualization should have no impact on performance for this task, since it only requires looking at the area around a single, highlighted element.

On task 3, there was again no significant difference between the three systems on the small dataset. However, with the large dataset, participants accuracy with EulerView was barely better than guessing (mean accuracy of approximately 15\%). Otherwise, MetroSets performed slightly better on the larger dataset than the smaller ($p=.046$, mean difference $\approx 15\%$).

All told, with respect to element based tasks (T1-T3), both LineSets and MetroSets maintained average accuracy of approximately 92\%. EulerView performed considerably worse, with an average accuracy of roughly 65\%.

On task 4, there was no significant difference between LineSets and MetroSets, regardless of the size of the dataset. However, EulerView performed significantly worse than either on both the large dataset ($p<.001$, mean difference $\approx 39\%$) and the small one ($p<.001,$ mean difference $\approx 61\%)$. Its performance on the small dataset was significantly worse than on the large ($p = .006$, mean difference $\approx 22\%$).

On task 5, there was no significant difference between the three systems on the small dataset. However, on the large dataset, both EulerView and LineSets performed very poorly, with average accuracy of roughly 14\%. MetroSets performed significantly better on the large dataset ($p < .001$, mean difference $\approx 56\%$). MetroSets' performance on the larger dataset was significantly worse than its performance on the smaller dataset, however ($p = .003$, mean difference $\approx 24\%$).

On task 6, MetroSets performed significantly better than EulerView on the small dataset ($p < .001,$ mean difference $\approx 36\%$). However, both EulerView and MetroSets performed significantly better than LineSets on the large dataset $(p < .001$, mean difference $\approx 53\%$).

Considering all set-based tasks (T4-T6), MetroSets maintained an average accuracy of approximately $85\%$. Meanwhile, LineSets had an average accuracy of roughly $64\%$, while EulerView's accuracy was only $50\%$.

The supplementary materials contains the mean and standard deviation for the accuracy of participants on each task, system, and dataset size.

\subsection{Time}

While we recorded the time taken by participants on each task, this data is difficult to analyze rigorously. As the study was online, we cannot confirm that our participants remained focused and attentive throughout the study. This fact, in combination with the general challenges associated with analyzing response time data (such as high skew and the tendency for high variance even when a single subject performs the same task multiple times) \cite{whelan2008effective}, led us to decide against detailed time analysis. Instead, we visualize the distribution of response times for each question, size, and system in \autoref{fig:time-violin}. In addition, descriptive statistics are given in the supplementary materials.

\begin{figure*}[t]
	\centering
	\subfloat {
		\includegraphics[width=0.33\textwidth]{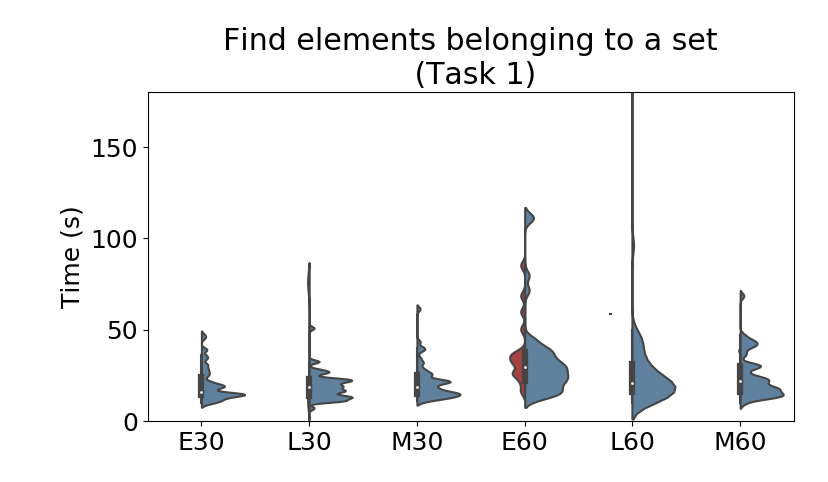}
	}
	\subfloat {
		\includegraphics[width=0.33\textwidth]{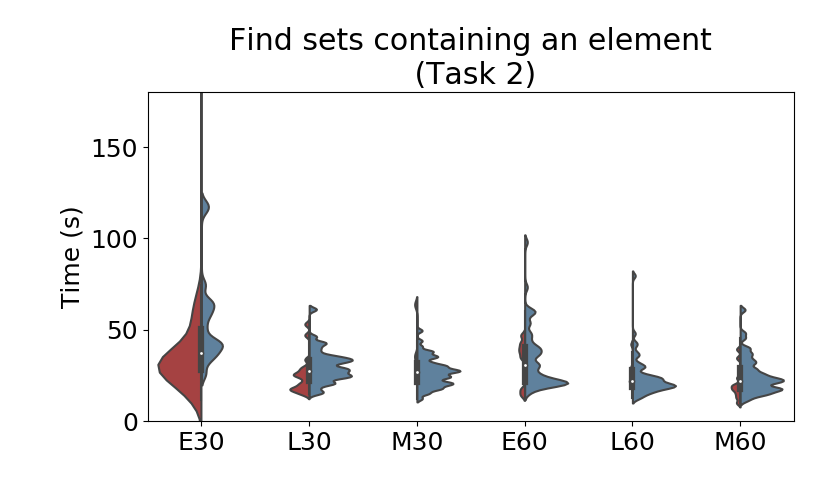}
	} 
	\subfloat {
		\includegraphics[width=0.33\textwidth]{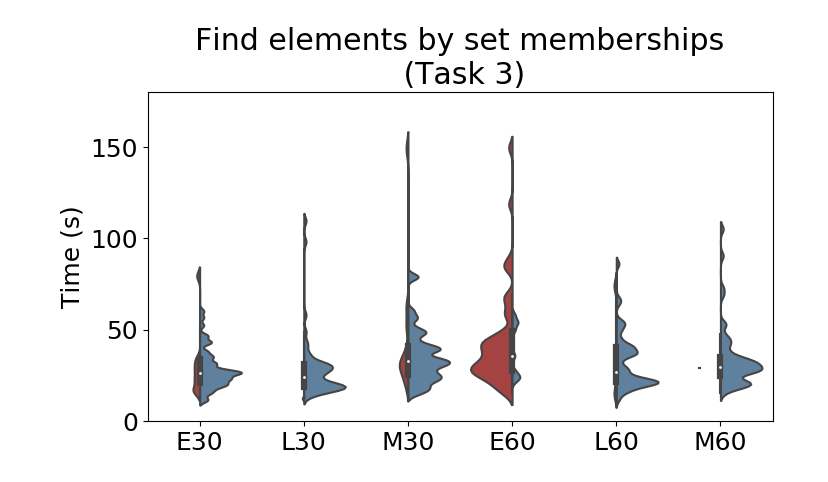}
	} \\
	\subfloat {
		\includegraphics[width=0.33\textwidth]{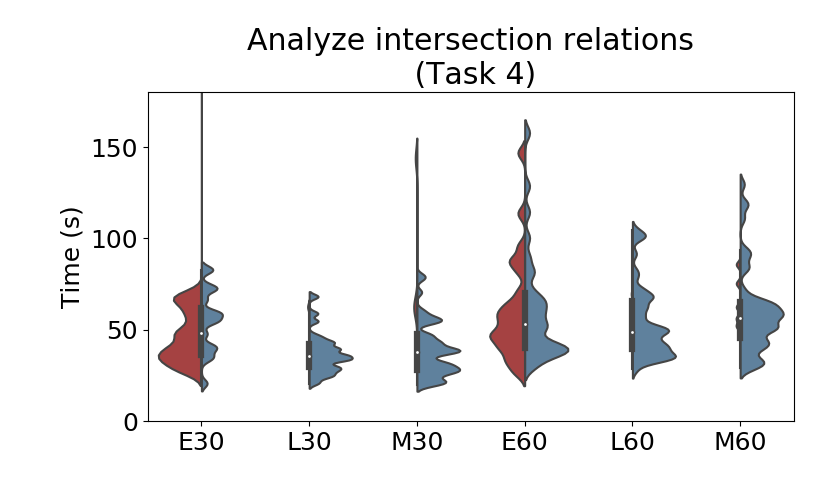}
	} 
	\subfloat {
		\includegraphics[width=0.33\textwidth]{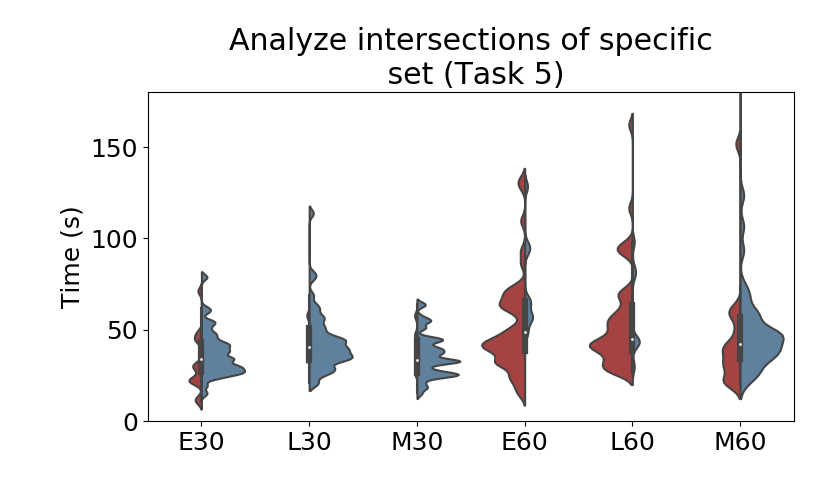}
	}
	\subfloat {
		\includegraphics[width=0.33\textwidth]{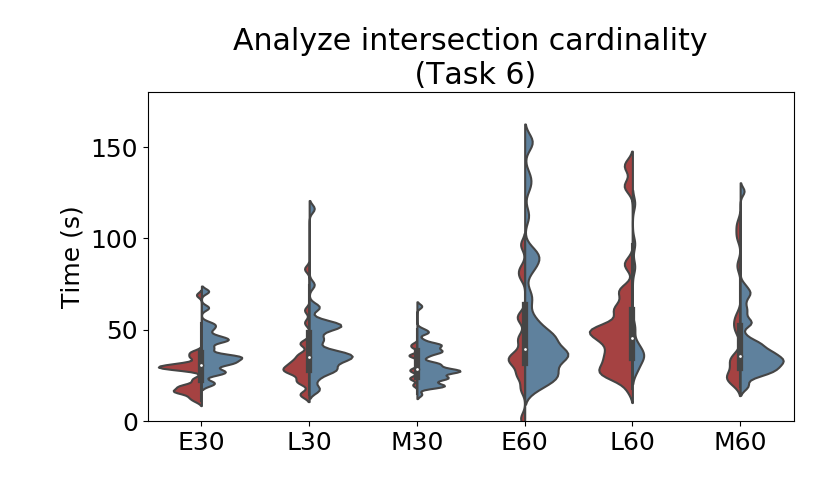}
	}
	\caption{Violin plots summarizing the distribution of times taken by participants on each system, size, and task. Each violin plot represents the distribution of times taken on that task with the corresponding system and size. The red region on the left represents the distribution of times taken on incorrect answers, while the blue region on the right represents the distribution of times for correct answers. The fatter a region is at a point, the more participants completed the task in roughly that much time. Note that a handful of extreme outliers are not visualized due to space constraints.}
	\label{fig:time-violin}
\end{figure*}

Comparing this data to our accuracy results in \autoref{fig:tukey-results} suggests that the variance in response time was greatest when participants struggled to find the right answer: see, e.g., the performance of EulerView on the small dataset for tasks T2 and T4. 
In general, however, there is no obvious pattern to the time taken as a function of the system. Each of the three techniques 
was the fastest for at least one task.

\subsection{Qualitative Feedback}
\begin{figure*}
  \centering
  \subfloat {
    \includegraphics[width=0.33\linewidth]{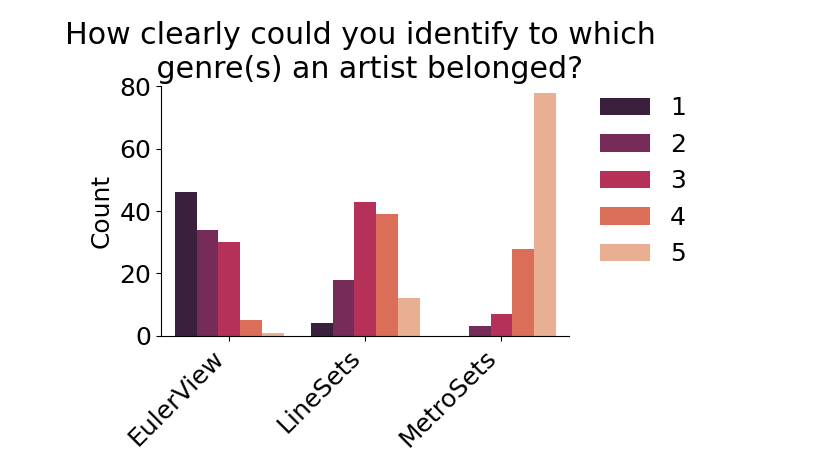}
  }
  \subfloat {
    \includegraphics[width=0.33\linewidth]{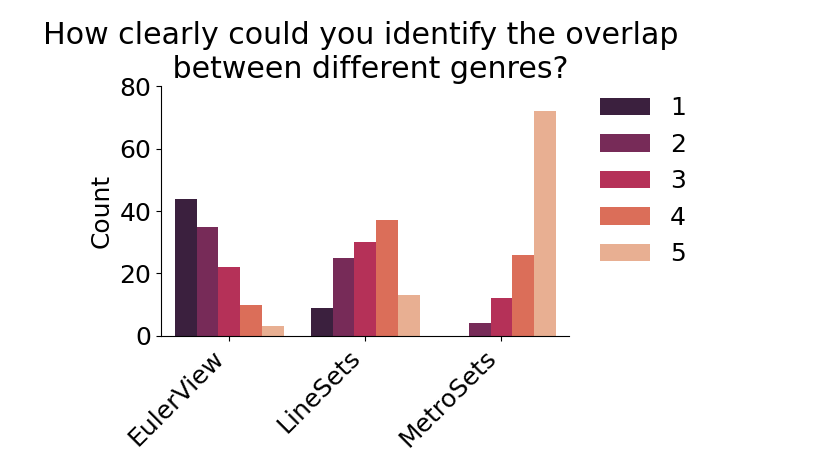}
  }
  \subfloat {
    \includegraphics[width=0.33\linewidth]{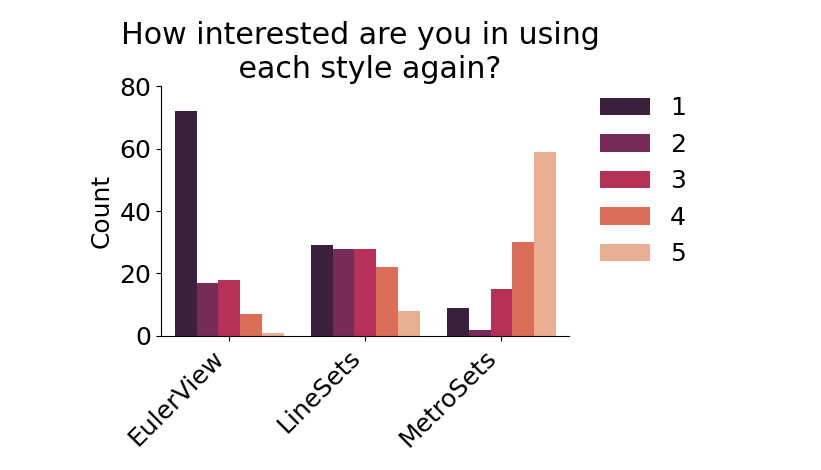}
  }
	\caption{Responses to post-questionnaire Likert scale questions. In question (a) and (b) the answer options ranged from 1 (poorly) to 5 (clearly). For question (c) the range was 1 (not) to 5 (very).}
	\label{fig:qual-results}
\end{figure*}

After completing the study, participants were invited to provide qualitative feedback on the systems and the study. This feedback took the form of four Likert-scale questions and a free-form text field. The distribution of answers for the Likert scale questions is presented in \autoref{fig:qual-results}.

We analyzed the results of the Likert scale questions using the Kruskal-Wallis test \cite{kruskal1952use}. This test showed that there was no significant difference in the frequency with which participants used pre-existing knowledge to answer questions for each system ($p=0.97$). We do find significant differences for the other questions asking about interest in the different visualization styles, ability to perform element-based tasks, and ability to perform set-based tasks ($p < 0.01$ in all three cases).

For each of the three questions with significant differences, we performed pairwise Mann-Whitney U tests\cite{mann1947test} between the systems, using Bonferroni correction\cite{bonferroni1936teoria}. This post-hoc analysis again revealed significant results ($p<0.01$ for all comparisons). We use the AUC measure for effect size\cite{hanley1982meaning}, which has the intuitive interpretation as the probability that a randomly chosen rating for one system will be larger than a randomly chosen rating for the second. We summarize the results below:

\begin{compactitem}
    \item For element-based tasks, the participants found it easier to identify the genres to which an artist belonged with MetroSets than with LineSets (AUC = 0.84) and with LineSets over EulerView (AUC = 0.82). The difference between MetroSets and EulerView was even greater (AUC = 0.97). 
    \item For set-based tasks, particpants found it easier to identify the overlap between genres in MetroSets than with LineSets (AUC = 0.8) and with LineSets over EulerView (AUC = 0.75). They also found MetroSets easier than EulerView (AUC = 0.92).
     \item The participants reported greater interest in using MetroSets than   LineSets (AUC = 0.8), and greater interest in using LineSets than EulerView (AUC = 0.7). They also preferred MetroSets to EulerView (AUC = 0.9).
\end{compactitem}

For the free-form response, we collected a total of 35 ($\sim60\%$) responses for the small dataset and 41 ($\sim71\%$) responses for the large dataset. 
The general opinion is that regions in EulerView are hard to distinguish, especially if the region depicts a intersection of more than two sets. The following response is a typical example: ``EulerView works great until you have more than 3 overlapping genres. For smaller sets of genres, I think EulerView would be preferable, but when artists belong to too many genres, it becomes difficult to identify all the patterns in the space.''

Participants had positive opinions about MetroSets and LineSets, with a slight preference for MetroSets when listing all three systems. 
Overlapping lines in LineSets was a recurring issue:
``Some instances of LineSets were very clear. Some other instances were unclear due to a lot of things going on in a very small region. I had the impression that such situations were avoided with MetroSets.''
This is in line with the observation we made in the statistical analysis that LineSets does not scale as well as MetroSets.
The color choice in all three systems was frequently critiqued. In particular, participants had difficulty distinguishing parallel lines with similar colors in MetroSets.
A list of all responses can be found in the supplementary material.

\subsection{Findings}
We hypothesized that, for both element and set based tasks, MetroSets would outperform LineSets and LineSets would outperform EulerView (H1 and H2). Our qualitative analysis supports these hypotheses: participants significantly preferred MetroSets over LineSets and LineSets over EulerView for both element-based and set-based tasks. 
Our quantitative analysis partially supports H1: MetroSets and LineSets both performed significantly better than EulerView on all element-based tasks, but there was no conclusive difference between MetroSets and LineSets.
Likewise, our data partially supports H2: on all of the set-based tasks, MetroSets outperformed either LineSets, EulerView, or both. However, there is no clear order between LineSets and EulerView: LineSets did much better on task T4, while EulerView did much better on task T6. The two performed comparably on task T5.

Finally, we hypothesized that MetroSets would scale better than LineSets or EulerView (H3). Our evidence supports this hypothesis: For all of the tasks except  T5, MetroSets' performance did not significantly degrade on the larger dataset, and the difference on task T5 was considerably larger for the other systems. Meanwhile, LineSets' performance degraded significantly on tasks T5 and T6, while EulerView's performance degraded significantly on tasks T1, T3, and T5.

We additionally performed exploratory analysis to determine if there were any meaningful correlations between demographic data, accuracy, and responses to the post-test questionnaire. 
In general, there is no obvious pattern between responses in one category and those in another. This is somewhat surprising: our demographic questions include factors such as familiarity with the visualization types, which intuitively could be relevant.

\section{Discussion}
\label{sec:limitations}

We next discuss some of the limitations of our study, as well as implications for improving existing set visualization approaches.

\subsection{Limitations}
While we aimed for a study design that allows to generalize the findings to visualizing set systems of similar characteristics, i.e., small to medium-sized data with less than 10--12 sets, less than 80--100 elements, and elements being members of 1--4 sets, we are aware of many limitations of our experiment. We list just a few examples below. 
We chose to use real-world music data to arouse the interest in voluntary participation in our study. This prevented us from having fine control over the visual stimuli; nonetheless, all three techniques used the same datasets and so the effect of the real-world data should apply to all. The use of real-world data also is associated with the risk that some participants might use prior knowledge when performing the tasks. However, the majority (84\%) of participants (self) reported never or almost never using pre-existing knowledge of music to help them answer questions. 

In the interest of keeping the study relatively short, we split the participants into two groups: one working with the small datasets and the other with the large datasets. Traditional confounding factors for between subjects studies likely apply here, hopefully ameliorated by the reasonably large number of participants in both settings.
As we decided to run the experiment online in a web browser rather than in a controlled lab setting, participants were using a variety of different screen sizes, input devices, and environmental conditions, which may have affected their task performance. We attempted to mitigate this issue by excluding experimental trials from smartphones or with screen resolutions below a minimum threshold. 

Some pairs of line colors could be difficult to discern and some of the textures used in EulerView were difficult to perceive than others. We attempted to mitigate this issue by using the same color scheme for both MetroSets and LineSets and by selecting tasks of similar difficulty levels across the different visualizations. However, it is possible that confounding factors associated with colors and textures do exists. 

We also acknowledge that our experiment only compares combinations of designs and implementations, and that the differences observed could be due to either. For example, in generating the LineSets stimuli, we contacted the original authors and followed their recommendation to use a standard force-based graph layout algorithm to determine the positions of elements. It is entirely possible that an implementation of LineSets using a different layout method would perform better. More generally, each of the designs we considered might admit better implementations than are currently available.

\subsection{Implications}

A recurring theme in the qualitative feedback was that participants had difficulty distinguishing colors in all three systems. The problem was most acute with EulerView, where regions often became muddled when many sets overlapped simultaneously. This difficulty may partly account for EulerView's comparatively poor performance in our experiment. However, given the strengths of EulerView relative to other Euler-based visualization systems, it may be worthwhile to revisit the technique, for example by using the smoother boundary contours proposed by Simonetto et al.~\cite{sas-sabied-16}. 

MetroSets and LineSets used the same color scheme in our experiment, and in both cases participants complained that some colors (particularly red and pink) were too difficult to distinguish. 
Usability of both systems 
would be improved by better use of colors. Borrowing from EulerView, it might be beneficial to explore using a smaller categorical color scheme supplemented by textures, such as dots or dashes along different lines.

For the most part, LineSets performed well in our study. However, it did occasionally produce ambiguous results, resulting in very low accuracy on some tasks. These problems are likely fixable. 
For example, representing elements with a glyph and then labeling, rather than representing elements directly with a label will likely help. From an algorithmic perspective, it may also be useful to compute curves collectively, rather than independently. In this way, it would be easy to detect overlapping lines and to split them apart. 

\section{Conclusions}
We evaluated three different systems for set visualization via a human-subjects experiment comparing their effectiveness both quantitatively and qualitatively. 
Our results include statistically significant differences, suggesting that MetroSets scales better and performs more consistently than EulerView or LineSets, and is also better-liked. We additionally considered implications for the design of all three systems.  

\ifCLASSOPTIONcompsoc
  \section*{Acknowledgments}
\else
  \section*{Acknowledgment}
\fi

The authors would like to thank all of the participants in our experiment, and especially the experts of our pilot study (Daniel Archambault, Helen Purchase, Silvia Miksch). 
This work is supported by NSF grants
CCF-1740858, CCF-1712119, and DMS-1839274 and by the Vienna
Science and Technology Fund (WWTF) through project ICT19-035.

\ifCLASSOPTIONcaptionsoff
  \newpage
\fi



\bibliographystyle{IEEEtran}
\bibliography{references-compact}

\begin{thebibliography}{10}
\providecommand{\url}[1]{#1}
\csname url@samestyle\endcsname
\providecommand{\newblock}{\relax}
\providecommand{\bibinfo}[2]{#2}
\providecommand{\BIBentrySTDinterwordspacing}{\spaceskip=0pt\relax}
\providecommand{\BIBentryALTinterwordstretchfactor}{4}
\providecommand{\BIBentryALTinterwordspacing}{\spaceskip=\fontdimen2\font plus
\BIBentryALTinterwordstretchfactor\fontdimen3\font minus
  \fontdimen4\font\relax}
\providecommand{\BIBforeignlanguage}[2]{{%
\expandafter\ifx\csname l@#1\endcsname\relax
\typeout{** WARNING: IEEEtran.bst: No hyphenation pattern has been}%
\typeout{** loaded for the language `#1'. Using the pattern for}%
\typeout{** the default language instead.}%
\else
\language=\csname l@#1\endcsname
\fi
#2}}
\providecommand{\BIBdecl}{\relax}
\BIBdecl

\bibitem{alsallakh2014visualizing}
B.~Alsallakh, L.~Micallef, W.~Aigner, H.~Hauser, S.~Miksch, and P.~J. Rodgers,
  ``The state-of-the-art of set visualization,'' \emph{Computer Graphics
  Forum}, vol.~35, no.~1, pp. 234--260, 2016.

\bibitem{mhsad-khvt-13}
W.~Meulemans, N.~Henry~Riche, B.~Speckmann, B.~Alper, and T.~Dwyer,
  ``Kelpfusion: A hybrid set visualization technique,'' \emph{{IEEE} Trans.
  Vis. Comput. Graph.}, vol.~19, no.~11, pp. 1846--1858, 2013.

\bibitem{csrmb-vsecdt-14}
P.~Chapman, G.~Stapleton, P.~Rodgers, L.~Micallef, and A.~Blake, ``Visualizing
  sets: An empirical comparison of diagram types,'' in \emph{Diagrammatic
  Representation and Inference (DIAGRAMS'14)}, ser. LNCS, vol. 8578.\hskip 1em
  plus 0.5em minus 0.4em\relax Springer, 2014, pp. 146--160.

\bibitem{ahrc-dslnvt-11}
B.~Alper, N.~Henry~Riche, G.~Ramos, and M.~Czerwinski, ``Design study of
  {LineSets}, a novel set visualization technique,'' \emph{{IEEE} Trans. Vis.
  Comput. Graph.}, vol.~17, no.~12, pp. 2259--2267, 2011.

\bibitem{DBLP:conf/diagrams/BaimagambetovS020}
A.~Baimagambetov, G.~Stapleton, A.~Blake, and J.~Howse, ``Evaluating
  visualizations of sets and networks that use {Euler} diagrams and graphs,''
  in \emph{Diagrammatic Representation and Inference (Diagrams'20)}, ser. LNCS,
  vol. 12169.\hskip 1em plus 0.5em minus 0.4em\relax Springer, 2020, pp.
  323--331.

\bibitem{DBLP:journals/isci/RodgersSAMBT16}
P.~J. Rodgers, G.~Stapleton, B.~Alsallakh, L.~Micallef, R.~Baker, and S.~J.
  Thompson, ``A task-based evaluation of combined set and network
  visualization,'' \emph{Information Sciences}, vol. 367-368, pp. 58--79, 2016.

\bibitem{lm-clmdv-19}
S.~Luz and M.~Masoodian, ``A comparison of linear and mosaic diagrams for set
  visualization,'' \emph{Information Visualization}, vol.~18, no.~3, 2019.

\bibitem{saa-favos-09}
P.~Simonetto, D.~Auber, and D.~Archambault, ``Fully automatic visualisation of
  overlapping sets,'' \emph{Computer Graphics Forum}, vol.~28, no.~3, pp.
  967--974, 2009.

\bibitem{jwkn-mvsmm-20}
B.~Jacobsen, M.~Wallinger, S.~Kobourov, and M.~Nöllenburg, ``Metrosets:
  Visualizing sets as metro maps,'' \emph{{IEEE} Trans. Vis. Comput. Graph.},
  vol.~27, no.~2, pp. 1257--1267, 2021.

\bibitem{rd-ued-10}
N.~Henry~Riche and T.~Dwyer, ``Untangling {Euler} diagrams,'' \emph{{IEEE}
  Trans. Vis. Comput. Graph.}, vol.~16, no.~6, pp. 1090--1099, 2010.

\bibitem{larsson2020eulerr}
\BIBentryALTinterwordspacing
J.~Larsson, \emph{{eulerr}: Area-Proportional {Euler} and {Venn} Diagrams with
  Ellipses}, 2020, {R} package version 6.1.0. [Online]. Available:
  \url{https://cran.r-project.org/package=eulerr}
\BIBentrySTDinterwordspacing

\bibitem{w-eaacved-12}
L.~Wilkinson, ``Exact and approximate area-proportional circular {Venn} and
  {Euler} diagrams,'' \emph{{IEEE} Trans. Vis. Comput. Graph.}, vol.~18, no.~2,
  pp. 321--331, 2012.

\bibitem{DBLP:conf/diagrams/RodgersFSH10}
P.~Rodgers, J.~Flower, G.~Stapleton, and J.~Howse, ``Drawing area-proportional
  {Venn}-3 diagrams with convex polygons,'' in \emph{Diagrammatic
  Representation and Inference (DIAGRAMS'10)}, ser. LNCS, vol. 6170.\hskip 1em
  plus 0.5em minus 0.4em\relax Springer, 2010, pp. 54--68.

\bibitem{sfrh-adedwc-12}
G.~Stapleton, J.~Flower, P.~J. Rodgers, and J.~Howse, ``Automatically drawing
  {E}uler diagrams with circles,'' \emph{J. Visual Languages and Computing},
  vol.~23, no.~3, pp. 163--193, 2012.

\bibitem{mr-efled-14}
L.~Micallef and P.~Rodgers, ``eulerforce: Force-directed layout for {Euler}
  diagrams,'' \emph{J. Visual Languages and Computing}, vol.~25, no.~6, pp.
  924--934, 2014.

\bibitem{sas-sabied-16}
P.~Simonetto, D.~W. Archambault, and C.~Scheidegger, ``A simple approach for
  boundary improvement of {Euler} diagrams,'' \emph{{IEEE} Trans. Vis. Comput.
  Graph.}, vol.~22, no.~1, pp. 678--687, 2016.

\bibitem{rzf-gedg-08}
P.~Rodgers, L.~Zhang, and A.~Fish, ``General {Euler} diagram generation,'' in
  \emph{Diagrammatic Representation and Inference (DIAGRAMS'08)}, ser. LNCS,
  vol. 5223, 2008, pp. 13--27.

\bibitem{srhz-iged-11}
G.~Stapleton, P.~Rodgers, J.~Howse, and L.~Zhang, ``Inductively generating
  {Euler} diagrams,'' \emph{{IEEE} Trans. Vis. Comput. Graph.}, vol.~17, no.~1,
  pp. 88--100, 2011.

\bibitem{cpc-bsrrwioev-09}
C.~Collins, G.~Penn, and S.~Carpendale, ``Bubble sets: Revealing set relations
  with isocontours over existing visualizations,'' \emph{{IEEE} Trans. Vis.
  Comput. Graph.}, vol.~15, no.~6, pp. 1009--1016, 2009.

\bibitem{dksw-kdpmv-12}
K.~Dinkla, M.~van Kreveld, B.~Speckmann, and M.~A. Westenberg, ``{Kelp}
  diagrams: Point set membership visualization,'' \emph{Computer Graphics
  Forum}, vol.~31, no.~3, pp. 875--884, 2012.

\bibitem{sgl-jsiativ-08}
J.~T. Stasko, C.~Görg, and Z.~Liu, ``Jigsaw: supporting investigative analysis
  through interactive visualization,'' \emph{Information Visualization},
  vol.~7, pp. 118--132, 2008.

\bibitem{drrd-pstfis-12}
M.~Dörk, N.~Henry~Riche, G.~A. Ramos, and S.~T. Dumais, ``Pivotpaths:
  Strolling through faceted information spaces,'' \emph{{IEEE} Trans. Vis.
  Comput. Graph.}, vol.~18, no.~12, pp. 2709--2718, 2012.

\bibitem{be-dhss-00}
F.~Bertault and P.~Eades, ``Drawing hypergraphs in the subset standard,'' in
  \emph{Graph Drawing (GD'00)}, ser. LNCS, vol. 1984.\hskip 1em plus 0.5em
  minus 0.4em\relax Springer, 2000, pp. 164--169.

\bibitem{aamh-rsivalos-13}
B.~Alsallakh, W.~Aigner, S.~Miksch, and H.~Hauser, ``Radial sets: Interactive
  visual analysis of large overlapping sets,'' \emph{{IEEE} Trans. Vis. Comput.
  Graph.}, vol.~19, no.~12, pp. 2496--2505, 2013.

\bibitem{jp-hpcdvd-87}
D.~S. Johnson and H.~O. Pollak, ``Hypergraph planarity and the complexity of
  drawing {V}enn diagrams,'' \emph{J. Graph Theory}, vol.~11, no.~3, pp.
  309--325, 1987.

\bibitem{bkmsv-psh-11a}
K.~Buchin, M.~van Kreveld, H.~Meijer, B.~Speckmann, and K.~Verbeek, ``On planar
  supports for hypergraphs,'' \emph{J. Graph Algorithms Appl.}, vol.~15, no.~4,
  pp. 533--549, 2011.

\bibitem{cgmny-spssh-19}
T.~Castermans, M.~van Garderen, W.~Meulemans, M.~Nöllenburg, and X.~Yuan,
  ``Short plane supports for spatial hypergraphs,'' \emph{J. Graph Algorithms
  Appl.}, vol.~23, no.~3, pp. 463--498, 2019.

\bibitem{ks-cmoct-03}
E.~Korach and M.~Stern, ``The clustering matroid and the optimal clustering
  tree,'' \emph{Mathematical Programming}, vol.~98, no. 1-3, pp. 385--414,
  2003.

\bibitem{kmn-mtshled-14}
B.~Klemz, T.~Mchedlidze, and M.~Nöllenburg, ``Minimum tree supports for
  hypergraphs and low-concurrency {E}uler diagrams,'' in \emph{Algorithm Theory
  (SWAT'14)}, ser. LNCS, vol. 8503.\hskip 1em plus 0.5em minus 0.4em\relax
  Springer, 2014, pp. 253--264.

\bibitem{bcps-psh-12}
U.~Brandes, S.~Cornelsen, B.~Pampel, and A.~Sallaberry, ``Path-based supports
  for hypergraphs,'' \emph{J. Discrete Algorithms}, vol.~14, pp. 248--261,
  2012.

\bibitem{kls-vcwpmd-07}
B.~H. Kim, B.~Lee, and J.~Seo, ``Visualizing set concordance with permutation
  matrices and fan diagrams,'' \emph{Interacting with Computers}, vol.~19, no.
  5-6, pp. 630--643, 2007.

\bibitem{smds-ovtlbd-14}
R.~Sadana, T.~Major, A.~D.~M. Dove, and J.~T. Stasko, ``Onset: {A}
  visualization technique for large-scale binary set data,'' \emph{{IEEE}
  Trans. Vis. Comput. Graph.}, vol.~20, no.~12, pp. 1993--2002, 2014.

\bibitem{lamy2019rainbio}
J.-B. Lamy and R.~Tsopra, ``{RainBio}: Proportional visualization of large sets
  in biology,'' \emph{IEEE Trans. Vis. Comput. Graph.}, 2019.

\bibitem{lex2014upset}
A.~Lex, N.~Gehlenborg, H.~Strobelt, R.~Vuillemot, and H.~Pfister, ``Upset:
  visualization of intersecting sets,'' \emph{IEEE Trans. Vis. Comput. Graph.},
  vol.~20, no.~12, pp. 1983--1992, 2014.

\bibitem{rsc-vswld-15}
P.~J. Rodgers, G.~Stapleton, and P.~Chapman, ``Visualizing sets with linear
  diagrams,'' \emph{{ACM} Trans. Comput. Hum. Interact.}, vol.~22, no.~6, pp.
  27:1--27:39, 2015.

\bibitem{stapleton2019efficacy}
G.~Stapleton, P.~Chapman, P.~Rodgers, A.~Touloumis, A.~Blake, and A.~Delaney,
  ``The efficacy of {Euler} diagrams and linear diagrams for visualizing set
  cardinality using proportions and numbers,'' \emph{PloS one}, vol.~14, no.~3,
  2019.

\bibitem{DBLP:journals/tvcg/FreilerMH08}
W.~Freiler, K.~Matkovic, and H.~Hauser, ``Interactive visual analysis of
  set-typed data,'' \emph{{IEEE} Trans. Vis. Comput. Graph.}, vol.~14, no.~6,
  pp. 1340--1347, 2008.

\bibitem{DBLP:journals/tvcg/KosaraBH06}
R.~Kosara, F.~Bendix, and H.~Hauser, ``Parallel sets: Interactive exploration
  and visual analysis of categorical data,'' \emph{{IEEE} Trans. Vis. Comput.
  Graph.}, vol.~12, no.~4, pp. 558--568, 2006.

\bibitem{DBLP:journals/tvcg/MicallefDF12}
L.~Micallef, P.~Dragicevic, and J.~Fekete, ``Assessing the effect of
  visualizations on bayesian reasoning through crowdsourcing,'' \emph{{IEEE}
  Trans. Vis. Comput. Graph.}, vol.~18, no.~12, pp. 2536--2545, 2012.

\bibitem{fruchterman1991graph}
T.~M. Fruchterman and E.~M. Reingold, ``Graph drawing by force-directed
  placement,'' \emph{Software: Practice and Experience}, vol.~21, no.~11, pp.
  1129--1164, 1991.

\bibitem{ellson2001graphviz}
J.~Ellson, E.~Gansner, E.~Koutsofios, S.~C. North, and G.~Woodhull, ``Graphviz
  -- open source graph drawing tools,'' in \emph{Graph Drawing (GD'01)}, ser.
  LNCS, vol. 2265.\hskip 1em plus 0.5em minus 0.4em\relax Springer, 2001, pp.
  483--484.

\bibitem{clark1924ishihara}
J.~Clark, ``The {Ishihara} test for color blindness.'' \emph{American Journal
  of Physiological Optics}, 1924.

\bibitem{fisher1992statistical}
R.~A. Fisher, ``Statistical methods for research workers,'' in
  \emph{Breakthroughs in Statistics: Methodology and Distribution}, S.~Kotz and
  N.~L. Johnson, Eds.\hskip 1em plus 0.5em minus 0.4em\relax Springer, 1992,
  pp. 66--70.

\bibitem{tukey1949comparing}
J.~W. Tukey, ``Comparing individual means in the analysis of variance,''
  \emph{Biometrics}, pp. 99--114, 1949.

\bibitem{whelan2008effective}
R.~Whelan, ``Effective analysis of reaction time data,'' \emph{The
  Psychological Record}, vol.~58, no.~3, pp. 475--482, 2008.

\bibitem{kruskal1952use}
W.~H. Kruskal and W.~A. Wallis, ``Use of ranks in one-criterion variance
  analysis,'' \emph{Journal of the American statistical Association}, vol.~47,
  no. 260, pp. 583--621, 1952.

\bibitem{mann1947test}
H.~B. Mann and D.~R. Whitney, ``On a test of whether one of two random
  variables is stochastically larger than the other,'' \emph{The annals of
  mathematical statistics}, pp. 50--60, 1947.

\bibitem{bonferroni1936teoria}
C.~Bonferroni, ``Teoria statistica delle classi e calcolo delle probabilita,''
  \emph{Pubblicazioni del R Istituto Superiore di Scienze Economiche e
  Commericiali di Firenze}, vol.~8, pp. 3--62, 1936.

\bibitem{hanley1982meaning}
J.~A. Hanley and B.~J. McNeil, ``The meaning and use of the area under a
  receiver operating characteristic (roc) curve.'' \emph{Radiology}, vol. 143,
  no.~1, pp. 29--36, 1982.

\bibitem{DBLP:conf/f-egc/AuberMMDLABPDM10}
D.~Auber, P.~Mary, M.~Mathiaut, J.~Dubois, A.~Lambert, D.~W. Archambault,
  R.~Bourqui, B.~Pinaud, M.~Delest, and G.~Melan{\c{c}}on, ``Tulip: a scalable
  graph visualization framework,'' in \emph{Extraction et gestion des
  connaissances (EGC'2010)}, vol. {RNTI-E-19}, 2010, pp. 623--624.

\bibitem{gansner2010gmap}
E.~R. Gansner, Y.~Hu, and S.~Kobourov, ``{GMap}: Visualizing graphs and
  clusters as maps,'' in \emph{Pacific Visualization Symposium
  (PacificVis'10)}.\hskip 1em plus 0.5em minus 0.4em\relax IEEE, 2010, pp.
  201--208.

\bibitem{hagberg2008exploring}
A.~Hagberg, P.~Swart, and D.~S~Chult, ``Exploring network structure, dynamics,
  and function using {NetworkX},'' Los Alamos National Lab., USA, Tech. Rep.
  LA-UR-08-5495, 2008.

\bibitem{o-mmw-03}
M.~Ovenden, \emph{Metro Maps of the World}.\hskip 1em plus 0.5em minus
  0.4em\relax Capital Transport Publishing, 2003.

\bibitem{nollenburg2014survey}
M.~Nöllenburg, ``A survey on automated metro map layout methods,'' in
  \emph{1st Schematic Mapping Workshop}, Essex, UK, 2014.

\bibitem{wntrn-stlfdmhp-20}
H.-Y. Wu, B.~Niedermann, S.~Takahashi, M.~J. Roberts, and M.~Nöllenburg, ``A
  survey on transit map layout -- from design, machine, and human
  perspectives,'' \emph{Computer Graphics Forum}, vol.~39, no.~3, pp. 619--646,
  2020.

\bibitem{niedermann2018algorithmic}
B.~Niedermann and J.-H. Haunert, ``An algorithmic framework for labeling
  network maps,'' \emph{Algorithmica}, vol.~80, no.~5, pp. 1493--1533, 2018.

\end{thebibliography}
%



%

\begin{IEEEbiography}[{\includegraphics[width=1in,height=1.25in,clip,keepaspectratio]{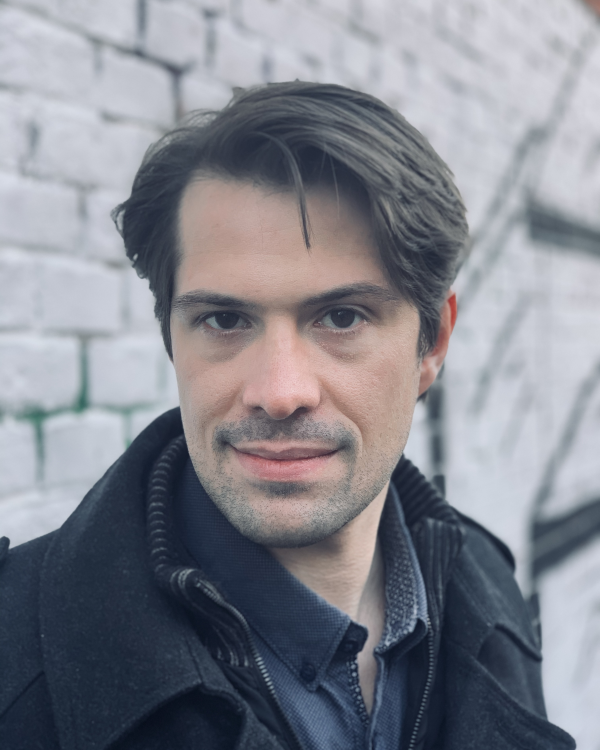}}]{Markus Wallinger} is a PhD student in the Algorithms and Complexity group of TU Wien. He received his BS degree in Computer Science from the University of Innsbruck and his MS degree in Visual Computing from TU Wien. His research focuses on algorithms for information visualization, graph theory and linear orders.
\end{IEEEbiography}
\begin{IEEEbiography}[{\includegraphics[width=1in,height=1.25in,clip,keepaspectratio]{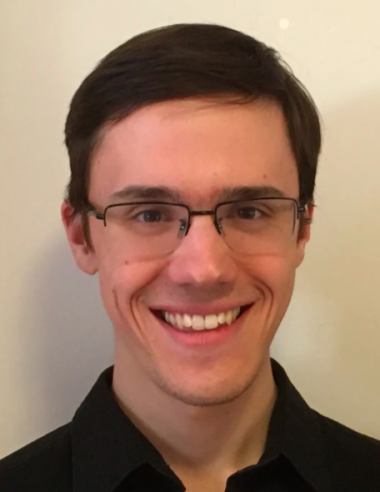}}]{Ben Jacobsen} is an undergraduate student at the University of Arizona majoring in Mathematics and Computer Science. His interests include graph theory, computer security, and the intersection of computer science, mathematics, and philosophy.
\end{IEEEbiography}
\begin{IEEEbiography}[{\includegraphics[width=1in,height=1.25in,clip,keepaspectratio]{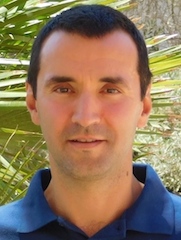}}]{Stephen Kobourov} is a Professor at the Department of Computer Science at the University of Arizona. He received a BS degree in Mathematics and Computer Science from Dartmouth College and MS and PhD degrees from Johns Hopkins University. His research interests include information visualisation, graph theory, and geometric algorithms.
\end{IEEEbiography}
\begin{IEEEbiography}[{\includegraphics[width=1in,height=1.25in,clip,keepaspectratio]{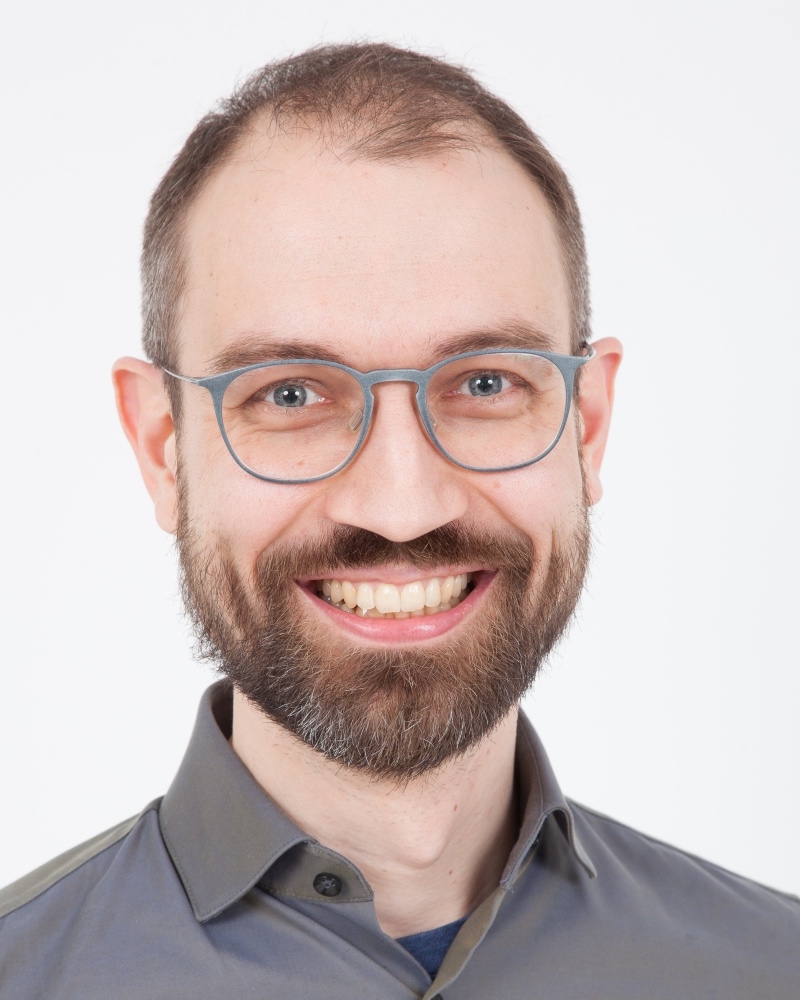}}]{Martin N\"ollenburg} is a Professor in the Algorithms and Complexity Group of TU Wien, Vienna, Austria. He received his PhD and habilitation degrees from Karlsruhe Institute of Technology (KIT), Germany, in 2009 and 2015, respectively. His  research  interests include graph drawing algorithms, computational geometry,  and information visualization  with  a focus on applications in network visualization and computational cartography.
\end{IEEEbiography}







\clearpage

\appendices

\section{Set Visualization Systems}
\subsection{EulerView}

EulerView by Simonetto et al.~\cite{saa-favos-09} generates Euler-like diagrams of set systems.
Euler diagrams consists of simple closed shapes in the two-dimensional plane that depict sets.
Each curve of the boundary of the shape divides the plane into two regions.
Elements inside the enclosed region are the members of the respective set, whereas elements outside the curve are not.
The EulerView algorithm extends Euler diagrams by allowing holes in regions or split regions, thus avoiding undrawable instances.


The implementation of EulerView is realized as a module in the Tulip~\cite{DBLP:conf/f-egc/AuberMMDLABPDM10} data visualization framework. 
Set elements are represented by white square glyphs with a black border and an attached label by default.  
Even though most attributes, such as glyph size, glyph shape, or label position, could be manipulated, we opted to keep the diagrams similar in style to the examples given in the original paper~\cite{saa-favos-09}, as any manipulation changed the rendering of textures drastically. The module does not incorporate the later improvements of smoother boundary contours proposed by Simonetto et al.~\cite{sas-sabied-16}. 
We were not able to use the EulerView module with a recent version of Tulip, even after contacting the original author. 
Eventually, we were able to use Tulip version 3.5 to create the output.

A major advantage of EulerView (and Euler-based systems generally) is that all elements with identical set membership are grouped together in the final visualization. It additionally solves several major issues plaguing Euler-based visualizations, such as the absence of element labels and the possibility of implying non-existent set overlaps.
Despite these strengths, there are also certain weaknesses inherent in this type of visualizations. In particular, the textures used to differentiate different regions  often become hard to distinguish when more than two or three sets overlap. This problem is exacerbated by the fact that the boundaries of different regions often overlap, making it difficult to identify them through their outlines.

\subsection{LineSets}

LineSets by Alper et al.~\cite{ahrc-dslnvt-11} is an overlay-based system for set visualization. It represents elements as points in the 2D plane and sets as continuous curves. If a set contains an element, then the corresponding curve passes through the corresponding point. 
Like other overlay-based systems, LineSets expects positions for each element 
as part of the input, and so the main algorithmic work consists of choosing the order in which each curve visits all elements that correspond to that set. Once such an order is determined, the curve is overlaid on top of the elements. 
Within these broad constraints, there are different ways to implement LineSets. We use the implementation provided in GMap\cite{gansner2010gmap}, which is available on github.

To apply LineSets to abstract data with no intrinsic spatial dimension, we follow the original designers of the system and create an initial embedding for the data using a force-based algorithm. Specifically, we treat the input as a graph, where two nodes are connected by an edge if they share at least one set, and generate a layout for it using the force-directed algorithm by Fruchterman-Reingold algorithm~\cite{fruchterman1991graph}, as implemented in NetworkX~\cite{hagberg2008exploring}. The positions generated for each node are then included in the input to LineSets.

It should be noted that, while this approach follows that taken by the authors of LineSets, it is not a fundamental component of the design. Alternative methods for determining layout, such as multi-dimensional scaling, would be equally viable, and the performance of LineSets in our study may to some extent reflect our particular choice of layout algorithm.


While LineSets visualizations are generally easy to interpret, two types of ambiguities can arise in some situations.
First, because labels are opaque and placed directly on top of the curves representing sets, it is sometimes difficult to determine whether a line actually goes through an element, or simply passes nearby. Second, lines travelling along the same sub-curve can overlap one another, making it hard to tell at a glance exactly how many lines connect a given pair of elements.

\subsection{MetroSets}

The core concept behind MetroSets~\cite{jwkn-mvsmm-20} is to apply the metro map metaphor to set visualization:
set elements are metro stations and different sets are different metro lines. 
Whenever a line passes through a station the associated element is considered a member of the set depicted by the line.
If an element is in multiple sets it will be depicted as an interchange station.
MetroSets creates a schematic drawing that adheres to typical metro map design rules~\cite{o-mmw-03,nollenburg2014survey,wntrn-stlfdmhp-20}: colored octolinear (horizontal, vertical, or 45$^\circ$-diagonals) lines, labeled stations, uniform  distance between stations, straight lines.

MetroSets uses a 4-step pipeline approach with two or more algorithms for each step that can be specified when creating a visualization. In our study, we 
use the default pipeline.
Elements are depicted by circles whose diameter is proportional to their set membership degree and lines are drawn as sequences of octolinear poly-lines between stations. 
Each line is given a distinct color (if less than 21 sets are present) from the Tableau 20 color palette.
Finally, each set element has a label assigned that is placed next to the corresponding station in $45^{\circ}$ steps and adheres to the design criteria for metro map labeling~\cite{niedermann2018algorithmic}.
The resulting drawing is rendered using the D3 javascript library\footnote{see \url{https://d3js.org/}} and a legend is added at the bottom right corner. 

While the visual style of MetroSets is similar to LineSets,  there are several important differences. In particular, MetroSets avoids issues with lines overlapping each other by adding regular white space between parallel lines. In addition, by placing labels near the elements, rather then on top of them, MetroSets reduces potential ambiguity when determining whether a line passes through a given element. 

\section{Additional statistical results}

This section contains further descriptive statistics and visualizations of the time and accuracy data we gathered through our experiments. \autoref{tab:desc-stats} presents the mean and standard deviation of accuracy for each task, size and system, while \autoref{tab:time_stats} does the same for time. Note that the mean and standard deviation are not robust against outliers, which are present in large number here because of the inherently skewed nature of response time data: Figure 5 visualizes the full distribution of response times for each task, size and system.


\begin{table}[h]
\caption{Accuracy of Participants}\label{tab:desc-stats}
\begin{tabularx}{\columnwidth}{c X|*{6}{X}}

 Task &&E30&L30&M30&E60&L60&M60\\
 \hline 
 T1&  mean  \newline  std & 0.97  \newline  0.18 & 0.97  \newline  0.18 & 1.00  \newline  0.00 & 0.74  \newline  0.44 & 0.98  \newline  0.13 & 0.95  \newline  0.22 \\
 \hline 
 T2&  mean  \newline  std & 0.33  \newline  0.47 & 0.69  \newline  0.47 & 0.95  \newline  0.22 & 0.88  \newline  0.33 & 0.93  \newline  0.26 & 0.83  \newline  0.38 \\
 \hline 
 T3&  mean  \newline  std & 0.86  \newline  0.35 & 0.97  \newline  0.18 & 0.83  \newline  0.38 & 0.16  \newline  0.37 & 0.95  \newline  0.22 & 0.98  \newline  0.13 \\
 \hline 
 T4&  mean  \newline  std & 0.33  \newline  0.47 & 0.97  \newline  0.18 & 0.93  \newline  0.26 & 0.55  \newline  0.50 & 0.95  \newline  0.22 & 0.91  \newline  0.28 \\
 \hline 
 T5&  mean  \newline  std & 0.78  \newline  0.42 & 0.95  \newline  0.22 & 0.95  \newline  0.22 & 0.16  \newline  0.37 & 0.14  \newline  0.35 & 0.71  \newline  0.46 \\
 \hline 
 T6&  mean  \newline  std & 0.48  \newline  0.50 & 0.62  \newline  0.49 & 0.84  \newline  0.37 & 0.72  \newline  0.45 & 0.21  \newline  0.41 & 0.76  \newline  0.43 \\
 \hline 
 \vspace{0.1em}
\end{tabularx}
\textit{This table summarizes statistics describing the accuracy of participants. Each row represents a single task, while each column represents a particular system and size.}
\end{table}

\begin{table}[h]
\caption{Timing of Participants}\label{tab:time_stats}
\begin{tabularx}{\columnwidth}{c X|*{6}{X}}

 Task &&E30&L30&M30&E60&L60&M60\\
 \hline 
 T1Time&  mean  \newline  std & 19.96  \newline  9.08 & 19.91  \newline  10.34 & 21.38  \newline  9.44 & 34.67  \newline  21.30 & 29.59  \newline  37.25 & 24.77  \newline  11.72 \\
 \hline 
 T2Time&  mean  \newline  std & 47.01  \newline  51.85 & 28.76  \newline  9.65 & 27.72  \newline  9.04 & 34.07  \newline  16.16 & 24.67  \newline  10.74 & 24.43  \newline  10.05 \\
 \hline 
 T3Time&  mean  \newline  std & 29.10  \newline  12.56 & 27.94  \newline  17.00 & 37.42  \newline  20.54 & 42.36  \newline  24.52 & 32.57  \newline  15.73 & 33.78  \newline  16.72 \\
 \hline 
 T4Time&  mean  \newline  std & 51.80  \newline  24.90 & 37.45  \newline  11.04 & 41.41  \newline  19.96 & 61.10  \newline  28.72 & 53.66  \newline  19.42 & 60.07  \newline  22.58 \\
 \hline 
 T5Time&  mean  \newline  std & 35.96  \newline  13.52 & 43.84  \newline  16.14 & 36.29  \newline  11.69 & 55.35  \newline  26.80 & 54.49  \newline  25.97 & 50.70  \newline  33.13 \\
 \hline 
 T6Time&  mean  \newline  std & 32.11  \newline  13.10 & 39.04  \newline  17.64 & 31.39  \newline  9.71 & 52.42  \newline  33.86 & 53.44  \newline  29.71 & 43.01  \newline  21.93 \\
 \hline 
 \vspace{0.1em}
\end{tabularx}
\textit{This table summarizes statistics describing the time taken on each task. Each row represents a single task, while each column represents a particular system and size.}
\end{table}

\end{document}